\documentclass[a4paper,fleqn,usenatbib]{mnras}

\usepackage{newtxtext,newtxmath}

\usepackage[T1]{fontenc}
\usepackage{ae,aecompl}

\usepackage{graphicx}	
\usepackage{amsmath}	
\usepackage{amssymb}	

\def\apj{ApJ}

\def\araa{ARA\&A}
\def\aap{A\&A}
\def\apjl{ApJL}
\def\apjs{ApJS}
\def\mnras{MNRAS}
\def\aj{AJ}
\def\nat{Nature}

\def\icarus{Icarus}

\title[WD0137-349B GCM]{Simplified 3D GCM modelling of the irradiated brown dwarf WD0137-349B}

\author[Lee et al.]{
Graham K. H. Lee$^{1}$\thanks{E-mail: graham.lee@physics.ox.ac.uk},
Sarah L. Casewell$^{2}$,
Katy L. Chubb$^{3}$,
Mark Hammond$^{1}$,
\newauthor
Xianyu Tan$^{1}$,
Shang-Min Tsai$^{1}$,
and Raymond T. Pierrehumbert$^{1}$
\\
$^{1}$Atmospheric, Oceanic \& Planetary Physics, Department of Physics, University of Oxford, Oxford OX1 3PU, UK \\
$^{2}$Department of Physics and Astronomy, University of Leicester, University Road, Leicester LE1 7RH, UK \\
$^{3}$SRON Netherlands Institute for Space Research, Sorbonnelaan 2, 3584 CA, Utrecht, Netherlands
}

\date{Accepted XXX. Received YYY; in original form ZZZ}

\pubyear{2020}

\begin{document}
\label{firstpage}
\pagerange{\pageref{firstpage}--\pageref{lastpage}}
\maketitle

\begin{abstract}
Context: White dwarf - Brown dwarf short period binaries (P$_{\rm orb}$ $\lesssim$ 2 hours) are some of the most extreme irradiated atmospheric environments known.
These systems offer an opportunity to explore theoretical and modelling efforts of irradiated atmospheres different to typical hot Jupiter systems. \\
Aims: We aim to investigate the three dimensional atmospheric structural and dynamical properties of the Brown dwarf WD0137-349B. \\
Methods: We use the three dimensional GCM model Exo-FMS, with a dual-band grey radiative-transfer scheme to model the atmosphere of WD0137-349B.
The results of the GCM model are post-processed using the three dimensional Monte Carlo radiative-transfer model \textsc{cmcrt}.\\
Results: Our results suggest inefficient day-night energy transport and a large day-night temperature contrast for WD0137-349B.
Multiple flow patterns are present, shifting energy asymmetrically eastward or westward depending on their zonal direction and latitude.
Regions of overturning are produced on the western terminator.
We are able to reproduce the start of the system near-IR emission excess at $\gtrsim$ 1.95 $\mu$m as observed by the GNIRS instrument.
Our model over predicts the IR phase curve fluxes by factors of $\approx$1-3, but generally fits the shape of the phase curves well.
Chemical kinetic modelling using \textsc{vulcan} suggests a highly ionised region at high altitudes can form on the dayside of the Brown dwarf. \\
Conclusions: We present a first attempt at simulating the atmosphere of a short period White dwarf - Brown dwarf binary in a 3D setting.
Further studies into the radiative and photochemical heating from the UV irradiation is required to more accurately capture the energy balance inside the Brown dwarf atmosphere.
Cloud formation may also play an important role in shaping the emission spectra of the Brown dwarf.
\end{abstract}

\begin{keywords}
stars: individual: WD0137-349B -- binaries: close -- stars: atmospheres -- planets and satellites: atmospheres -- brown dwarfs -- radiative transfer
\end{keywords}



\section{Introduction}

Currently only a few post-common envelope, short period White dwarf - Brown dwarf (henceforth WD-BD) binary systems have been detected:
GD1400 \citep{Farihi2004,Dobbie2005,Burleigh2011}, WD0137-349 \citep{Maxted2006, Burleigh2006}, WD0837+185 \citep{Casewell2012b}, NLTT 5306 \citep{Steele2013}, SDSS J141126.20+200911.1 \citep{Beuermann2013, Littlefair2014, Casewell2018b}, SDSS J155720.77+091624.6 \citep{Farihi2017}, SDSS J1205-0242 \citep{Parsons2017, Rappaport2017}, SDSS J1231+0041 \citep{Parsons2017} and EPIC212235321 \citep{Casewell2018}.
Despite their rarity, with estimates of a $\approx$ 0.5\% rate of BD companions to WDs \citep{Steele2011}, these systems offer a unique insight into the properties of irradiated atmospheres in more extreme conditions than typical hot Jupiter systems.

The BD companion to WD0137-349 was first inferred by \citet{Maxted2006} through high-resolution radial velocity measurements, finding a mass ratio of $\approx$0.134 for the system.
Fitting the spectral data with a WD atmospheric model yielded a mass of $\sim$0.39 M$_{\rm \odot}$, placing the companion in the BD mass regime at $\sim$53 M$_{\rm J}$.
A near infrared (near-IR) excess was also hinted at in archival 2MASS \citep{Skrutskie2006} photometric data.
Further observations using the Gemini GNIRS instrument by \citet{Burleigh2006} confirmed the near-IR excess beyond $\approx$1.95 $\mu$m, providing a direct detection of the BD companion thermal emission.
\citet{Casewell2015} performed a comprehensive observational campaign spanning the \textit{V}, \textit{R}, \textit{I}, \textit{J}, \textit{H} and \textit{K$_{s}$} bands and also obtained \textit{Spitzer} data for the 3.6, 4.5, 5.8 and 8$\mu$m photometric bands.
They presented phase curves for the BD companion, and calculated a day-night temperature contrast of $\gtrsim$ 500 K in most of the IR bands.
\citet{Longstaff2017} presented spectroscopic detections of H$_{\alpha}$, He, Na, Mg, Si, K, Ca, Ti and Fe emission from the BD companion, suggesting molecular dissociation occurring in the upper atmosphere of the BD.

Studying WD-BD binaries present an opportunity to explore the nature of irradiated atmospheres in more `extreme' conditions than typical hot Jupiter (HJ) systems.
Atmospheric modelling of the BD is a challenging prospect due to several factors:
\begin{enumerate}
\item Moderate irradiation from the WD (T$_{\rm eq}$ $\approx$ 1000-2000 K), with $>5$ $\%$ of the stellar flux occurring at UV wavelengths.
\item High surface gravity (g $\gtrsim$ 1000 m s$^{-2}$).
\item Fast rotational speeds (P$_{\rm orb}$ $\lesssim$ 120 mins), assuming tidal locking.
\end{enumerate}
Even when taken individually, these factors represent a significant regime change from typical HJ conditions.
Examining these systems is therefore a test of current theories and models in a new context, and to provide the community with a holistic understanding of irradiated atmospheres.

In this initial study, we model the atmosphere of the companion BD in the WD0137-349 system.
We perform 3D global circulation models (GCMs) of the BD atmosphere with a simplified two-band grey radiative-transfer scheme.
The thermal structure of the GCM is then post-processed using a 3D radiative-transfer code and compared to the observational data from \citet{Burleigh2006} and \citet{Casewell2015}.
In Section \ref{sec:previous_modelling}, we briefly review current atmospheric modelling efforts of WD-BD binaries and the dynamical expectations from previous HJ studies.
Section \ref{sec:GCM_modelling} presents details of our GCM simulation and adopted parameters.
Section \ref{sec:dynamics} presents the results of our GCM simulation.
Section \ref{sec:post-processing} presents post-processing of our GCM simulation and comparison to available observational data.
Section \ref{sec:discussion} presents the discussion of our results and Section \ref{sec:conclusions} contains the summary and conclusions.

\section{Previous WD-BD modelling}
\label{sec:previous_modelling}

To date, modelling efforts for WD-BD systems have been rare in the literature.
1D radiative-convective modelling of WD0137-349B performed in \citet{Casewell2015} suggest that the photometry of the BD is best fit with a full circulation efficiency, and without the presence of strong optical wavelength opacity sources such as TiO and VO molecules.
UV photochemical effects such as H$_{2}$ fluorescence and H$_{3}^{+}$ formation and emission were also examined as candidates for boosting the \textit{K$_{s}$} band emission flux.
Similar modelling and conclusions were found for the SDSS J141126.20+200911.1 system in \citet{Casewell2018b}.
\citet{Longstaff2017} adapted a \textsc{drift-phoenix} \citep{Witte2009, Witte2011} (T$_{\rm eff}$ = 2000 K, log g = 5, [M/H] = 0) atmospheric profile with a hot chromospheric region to examine the thermal dissociation and ionisation profiles of the species detected in their observations.

\citet{HernandezSantisteban2016} use an energy balance model with a simplified redistribution efficiency parameter for the WD-BD interacting binary system SDSS J143317.78+101123.3.
Their best fit parameters suggest poor day/night energy transport efficiency.

\subsection{Dynamical expectations from HJ studies}
\label{sec:expectations}

WD-BD short period binaries inhabit a unique parameter regime, namely moderate to strong irradiation with a fast rotation rate.
Table \ref{tab:parameters} shows our adopted WD0137-349 system parameters.

\citet{Komacek2016,Komacek2017} and \citet{Komacek2018} examine the effect of increasing irradiation on hot Jupiter atmospheric circulation show that with increasing effective temperature, the radiative timescales become shorter, resulting in a higher day-night temperature contrast and inefficient day-night energy transport.
Several studies have examined the effects of rotation rate on the dynamical regime of the atmosphere, with and without the assumption of tidal locking for example, \citet{Showman2008,Showman2009,Kataria2013,Rauscher2014,Showman2015,Komacek2017} and \citet{Penn2017}.
In the short orbital period and forcing regime of WD0137-349B, the above studies suggest the formation of a Matsuno-Gill flow pattern \citep{Matsuno1966,Gill1980} commonly seen in HJ simulations, along with a multiple banded jet structure due to the higher rotation rate.

\citet{Tan2019} examined the effects of both increasing irradiation and rotation rates in the context of modelling ultra hot Jupiter atmospheres, finding similar conclusions to the studies above without H$_{2}$ dissociation and recombination.
Including the cooling/heating effects of H$_{2}$ dissociation/recombination reduced the day-night temperature contrasts in their simulations compared to no H$_{2}$ dissociation/recombination.

An estimate for the radiative-timescale, $\tau_{\rm rad}$ [s], is given by \citep{Showman2002}
\begin{equation}
 \tau_{\rm rad} \sim \frac{p}{g}\frac{c_{p}}{4\sigma T^{3}},
\end{equation}
where c$_{p}$ [J kg$^{-1}$ K$^{-1}$] is the heat capacity at constant pressure. For WD0137-349B, taking p = 10 bar, g = 1000 m s$^{-2}$, c$_{p}$ = 14308 J kg$^{-1}$ K$^{-1}$ and T = T$_{\rm eq}$ = 1995 K gives $\tau_{\rm rad}$ $\sim$ 7945 s.
This value is small compared to typical values at this pressure ($\tau_{\rm rad}$ $\sim$ 10$^{6}$) in HJ atmospheres \citep[e.g.][]{Showman2008}, and is more typical of mbar pressures in HJ atmospheres.
This suggests the high gravity has a major effect reducing the heat redistribution efficiency by lowering the radiative-timescales as a whole in the atmosphere.

We also examine derived atmospheric regime parameters similar to \citet{Kataria2016}.
The Rossby number, Ro  is given by
\begin{equation}
Ro = \frac{U}{fL},
\end{equation}
where f = 2$\Omega\sin\phi$, U a characteristic horizontal velocity, which we follow \citet{Kataria2016} and approximate as the global rms velocity expression from \citet{Lewis2010} at the IR photospheric pressure (10 bar) yielding U $\approx$ 1000 m s$^{-1}$.
We calculate f at mid-latitude and assume L = 1.1 R$_{\rm jup}$.
We estimate the Rhines scale, L$_{\beta}$ [m], from
\begin{equation}
 L_{\beta} = \pi \sqrt{\frac{U}{\beta}},
\end{equation}
where $\beta$ = 2$\Omega\cos\phi$/R$_{p}$ is evaluated at the equator.
The Rossby deformation radius, L$_{\rm D}$ [m], is estimated through
\begin{equation}
L_{\rm D} = \frac{NH}{f},
\end{equation}
where N is the Brunt-V\"{a}is\"{a}l\"{a} frequency and f calculated at mid-latitude.
Table \ref{tab:regime} presents these values for WD0137-349B along with a selection of other objects.

From these estimates, WD0137-349B occupies a distinct dynamical regime.
It is most like Jupiter with small Ro, L$_{\rm D}$ and L$_{\beta}$, but occupies a radiative regime more typical of HJs.

\section{GCM modelling using exo-FMS}
\label{sec:GCM_modelling}

\begin{table}
\centering
\caption{Adopted physical parameters and derived characteristics of the WD0137-349 system following \citet{Maxted2006, Burleigh2006, Casewell2015}.
$^{*}$ denotes an estimated value}
\begin{tabular}{c c c c c c c c}  \hline \hline
 T$_{\rm eff, WD}$ & R$_{\rm WD}$  &  M$_{\rm BD}$  & R$_{\rm BD}$   & a  & P$_{\rm orb}$ & inc. & dist.   \\ \relax
 [K] & [R$_{\rm \odot}$] & [M$_{\rm J}$] &  [R$_{\rm J}$] & [R$_{\rm \odot}$] & [min]  & [$^{\circ}$] & [pc] \\ \hline
 16500  & 0.019 & 53 & 1.1$^{*}$ & 0.65 & 116  & 35 & 102 \\
 \hline
\end{tabular}
\label{tab:parameters}
\end{table}

\begin{table*}
\centering
\caption{Characteristic values and scales of WD0137-349B compared to a selection of other planets.
Values were sourced from \citet{Parmentier_thesis}.
$^{*}$ denotes an estimated value.
Kelt-9b, WASP-43b and WASP-121b were estimated following the prescription in \citet{Parmentier_thesis}, where the characteristic velocity is taken as a range between 100 and 1000 m s$^{-1}$.}
\begin{tabular}{c c c c c c c c c}  \hline \hline
 Object & R$_{\rm p}$ & $\Omega$ & g & T$_{\rm eq}$  & H & Ro  & L$_{\rm D}$ & L$_{\beta}$  \\ \relax
        & [R$_{\rm J}$] & [rad s$^{-1}$] &[m s$^{-2}$] & [K] &  [km] & [-]  & [R$_{\rm p}$] & [R$_{\rm p}$] \\ \hline
   WD0137-349B & 1.1$^{*}$ &  9.155 $\times$ 10$^{-4}$ &  1086$^{*}$  & 1995 & 6.64 & $\approx$0.01   & $\approx$0.01  & $\approx$0.26 \\
   Jupiter & 1.0 & 1.4 $\times$ 10$^{-4}$ & 23.1 & 124 & 20 & 0.02 & 0.03 & 0.1 \\
   HD 209458b & 1.36 & 2.1 $\times$ 10$^{-5}$ & 10.2 & 1450 & 520 & 0.04-1.0 & 0.4 & 0.5-3 \\
   Kelt-9b & 1.89 & 4.91 $\times$ 10$^{-5}$ & 20.0 & 4051 & 734 & 0.01-0.4 & 0.2 & 0.3-2 \\
   WASP-43b & 1.04 & 8.94 $\times$ 10$^{-5}$ & 47.4 & 1441 & 110 & 0.01-0.4 & 0.1 & 0.3-2 \\
   WASP-121b & 1.87 & 5.70 $\times$ 10$^{-5}$ & 8.4 & 2358 & 1010 & 0.01-0.4 & 0.1 & 0.3-2 \\
 \hline
\end{tabular}
\label{tab:regime}
\end{table*}

\begin{table*}
\centering
\caption{Adopted GCM simulation parameters. Adapted for the WD0137-349 system from the \citet{Heng2011b} HJ simulation parameters.}
\begin{tabular}{c c c l}  \hline \hline
 Symbol & Value & Unit & Description \\ \hline
 F$_{\rm 0}$ & 3.59 $\times$ 10$^{6}$ & W m$^{-2}$ & Stellar irradiation constant \\
 A$_{\rm B}$ & 0.1 & - & Bond albedo \\
 T$_{\rm int}$ & 500 & K & Internal temperature \\
 P$_{\rm 0}$ & 220 & bar & Reference surface pressure \\
 $\tau_{\rm S_{\rm 0}}$ & 15.68 & - & Shortwave surface optical depth \\
 $\tau_{\rm L_{\rm eq}}$ & 22.0 & - & Longwave surface optical depth \\
 n$_{\rm S}$ &  1 & - & Shortwave power-law index \\
 n$_{\rm L}$ &  1 & - & Longwave power-law index \\
 c$_{\rm P}$ & 14308.4 & J K$^{-1}$ kg$^{-1}$ & Specific heat capacity \\
 R & 4593 &  J K$^{-1}$ kg$^{-1}$  & Ideal gas constant \\
 $\kappa$ & 0.321 & J K$^{-1}$ kg$^{-1}$  & Adiabatic coefficient \\
 g$_{\rm BD}$ & 1000 & m s$^{-2}$ & Acceleration from gravity \\
 R$_{\rm BD}$ & 7.86 $\times$ 10$^{4}$ & km & Radius of Brown dwarf \\
 $\Omega_{\rm BD}$ & 9.155 $\times$ 10$^{-4}$ & rad s$^{-1}$ & Rotation rate of Brown dwarf \\
 $\Delta$t & 20 & s & Simulation time-step \\
 T$_{\rm init}$ & 1824 & K & Initial isothermal temperature \\
 N$_{\rm v}$ & 50 & - & Vertical resolution \\
 d$_{\rm 2}$ & 0.02 & - & div. dampening coefficient \\
\hline
\end{tabular}
\label{tab:GCM_parameters}
\end{table*}

\begin{figure} 
   \centering
   \includegraphics[width=0.49\textwidth]{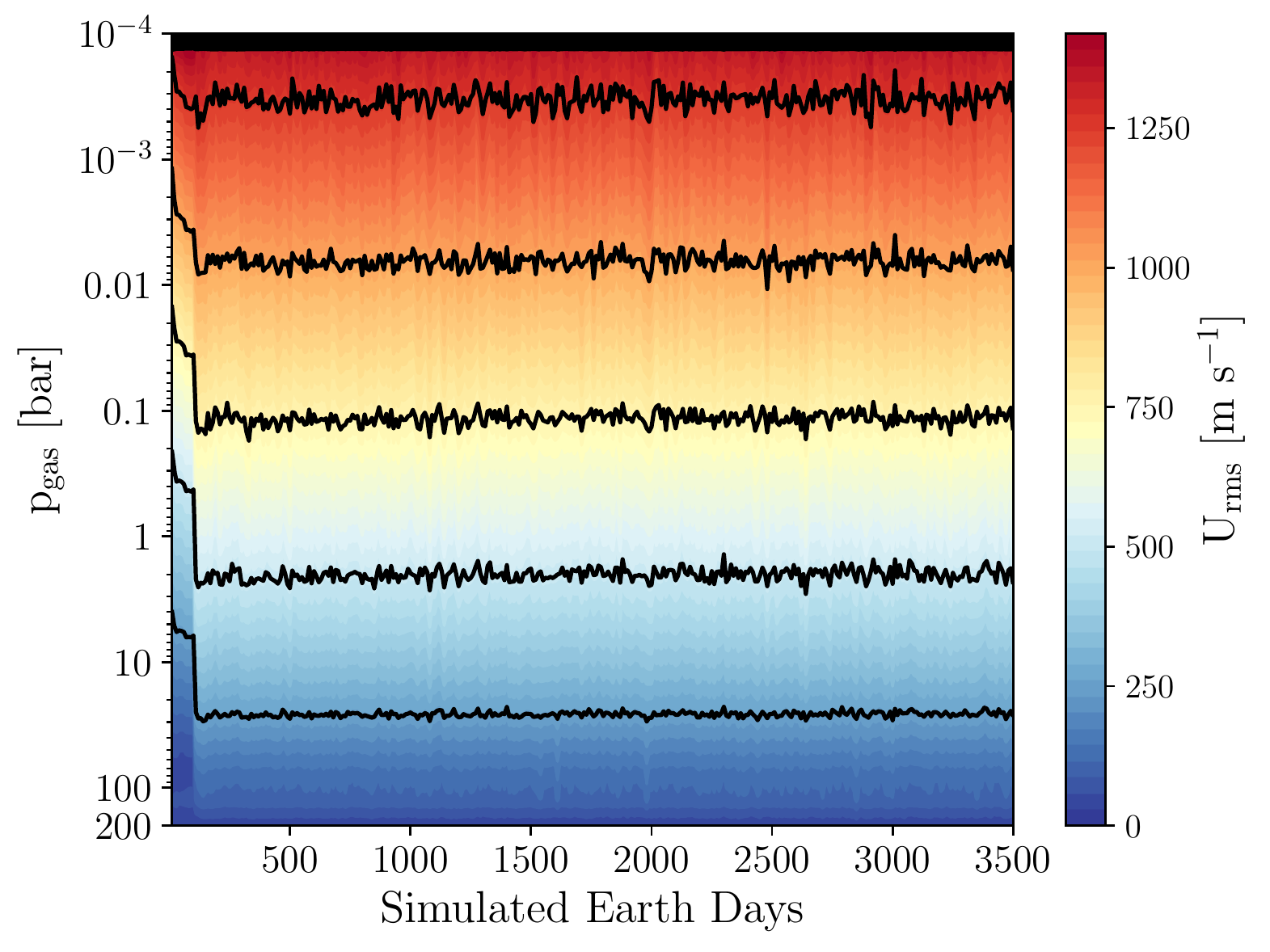}
   \caption{Global rms velocity as function of pressure with simulated time.
   The jump at 100 days corresponds to the turning off of a stronger Rayleigh drag used to stabilise the simulation during spin up.}
   \label{fig:RMS}
\end{figure}

We use the three dimensional, finite-volume Flexible Modelling System (FMS) GCM model \citep{Lin2004}, previously used to model terrestrial exoplanet atmospheres (Exo-FMS) \citep{Pierrehumbert2016, Hammond2017}.
We update Exo-FMS to use a cubed-sphere grid \citep[e.g.][]{Showman2009} with a resolution of C48 ($\approx$ 192 longitude $\times$ 96 latitude).
This set up has recently been benchmarked for hot Jupiter-like conditions (Lee et al. in prep.).

Exo-FMS evolves the primitive equations of meteorology \citep[e.g.][]{Mayne2014, Komacek2016} with a convective adjustment scheme.
We adopt a similar GCM setup to the dual band hot Jupiter simulations performed in \citet{Heng2011b}, with the appropriate conditions for WD0137-349B.
An assumed radius of 1.1 Jupiter radii (R$_{\rm J}$ = 7.1492 $\cdot$ 10$^{4}$ km), surface gravity of g$_{\rm BD}$ = 1000 m s$^{-2}$ and Bond albedo of A$_{\rm B}$ = 0.1 is taken for the BD.
We use a 50 vertical layer set up which is set using a hybrid sigma coordinate grid, approximately log spaced in pressure between 220 and 10$^{-4}$ bar.
A summary of the input parameters used for the GCM model is given in Table \ref{tab:GCM_parameters}.

The model is run for a total of 3500 simulated Earth days, equivalent to $\approx$ 43448 orbital periods.
The outputs presented here are an average of the last 500 days of simulation.
In Fig. \ref{fig:RMS} we show the global rms velocity at each pressure level during the 3500 day simulation.

\subsection{Radiative Transfer}

In order to avoid the difficulties associated with modelling UV radiative heating and associated photochemical heating, for radiative-transfer inside the GCM we use a double-grey scheme.
We assume the infrared and optical grey opacity values from \citet{Guillot2010} given by $\kappa_{\rm IR}$ = 10$^{-2}$ cm$^{2}$ g$^{-1}$ and  $\kappa_{\rm V}$ = 6 $\times$ 10$^{-3}$$\sqrt{(T_{\rm irr}/2000)}$ = 7.134 $\times$ 10$^{-3}$ cm$^{2}$ g$^{-1}$.
Due to the higher surface gravity of the BD (g$_{\rm BD}$ $\sim$ 1000 m s$^{-2}$) compared to typical hot Jupiters (g$_{\rm HJ}$ $\sim$ 10 m s$^{-2}$), the atmospheric vertical extension is much reduced compared to a hot Jupiter.
This results in the infrared optical depth at the reference pressure of our model (220 bar) of $\tau_{\rm L_{\rm eq}}$ = 22 and visual optical depth of $\tau_{\rm S_{\rm 0}}$ = 15.68, substantially lower than typical HJ simulations \citep[e.g.][]{Heng2011b, Rauscher2012}.

With the absence of a deep optically thick region, this suggests the dayside to be mostly dominated by the irradiation from the white dwarf.
The radiative timescale on the dayside is estimated to be short for the WD0137-349B parameters \citep[e.g.][]{Showman2008}, suggesting the dayside profiles are expected to be near radiative equilibrium.
Since day-night energy redistribution by flows is suggested to be weak for such systems (Sect. \ref{sec:expectations}), nightside profiles are expected to be colder and primarily controlled by the internal flux.

To estimate the internal flux, we assume WD0137-349B follows the HJ population trends and use the expression of \citet{Thorngren2019}.
This yields a value of T$_{\rm int}$ = 665 K, we therefore adopt a T$_{\rm int}$ of 500 K for this study as a more tractable value in our GCM model.

\subsection{Numerical Stability}

To aid numerical stability of the GCM we apply the `basal' drag formulation of \citet{Liu2013} in the lower atmospheric regions, commonly used in (ultra) hot Jupiter GCM studies \citep[e.g.][]{Komacek2016, Tan2019, Carone2019} motivated as a mimic to magnetic drag forces.
This takes the form of a pressure dependent linear drag, $F_{\rm dr}(p)$ [m s$^{-2}$], in the horizontal momentum equation \citep{Komacek2016, Carone2019}
\begin{equation}
F_{\rm dr}(p) = -\frac{\textbf{\textit{v}}}{\tau_{\rm dr}(p)},
\end{equation}
where \textbf{\textit{v}} [m s$^{-1}$] is the local velocity vector and $\tau_{\rm dr}(p)$ [s] the pressure dependent drag timescale.
$\tau_{\rm dr}(p)$ is given as a linear function of pressure between a prescribed top and bottom pressure level (p$_{\rm dr,t}$ and p$_{\rm dr,b}$ respectively) where the drag force is present.
\begin{equation}
\tau_{\rm dr}(p) = \tau_{\rm dr,b} \frac{(p - p_{\rm dr,t})}{(p_{\rm dr,b} - p_{\rm dr,t})},
\end{equation}
where $\tau_{\rm dr,b}$ [s] is the drag timescale at the simulation lower boundary (here taken as 1 Earth day).
We take p$_{\rm dr,b}$ to be the lower boundary pressure (220 bar) and p$_{\rm dr,t}$ = 10 bar.

\section{Atmospheric structure of WD0137-349B}
\label{sec:dynamics}

\begin{figure} 
   \centering
   \includegraphics[width=0.49\textwidth]{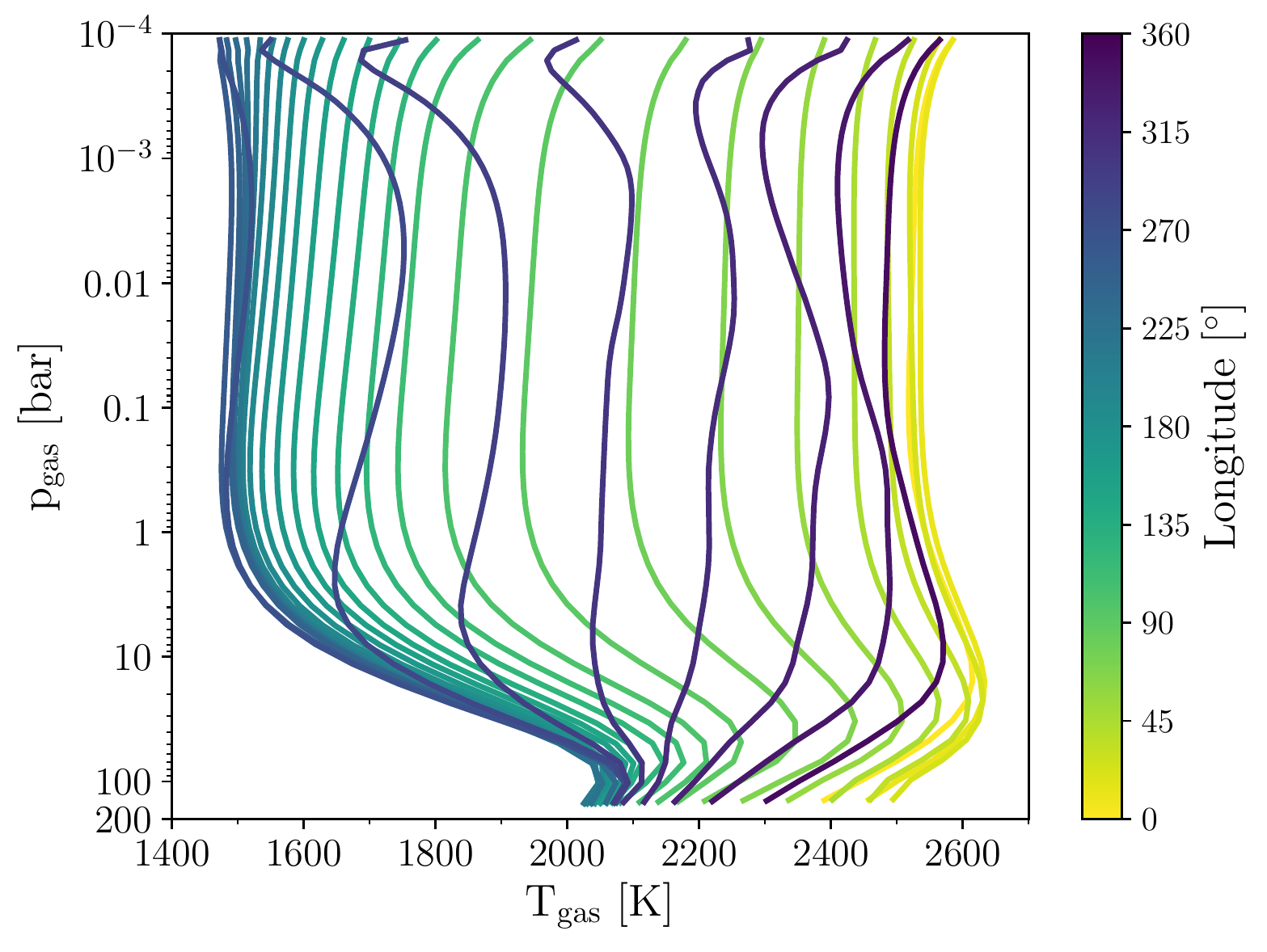}
   \caption{One dimensional temperature-pressure profiles at the equatorial region of the GCM output.
   The colour bar shows the longitude of the profile.}
   \label{fig:Tp}
\end{figure}

\begin{figure*} 
   \centering
   \includegraphics[width=0.49\textwidth]{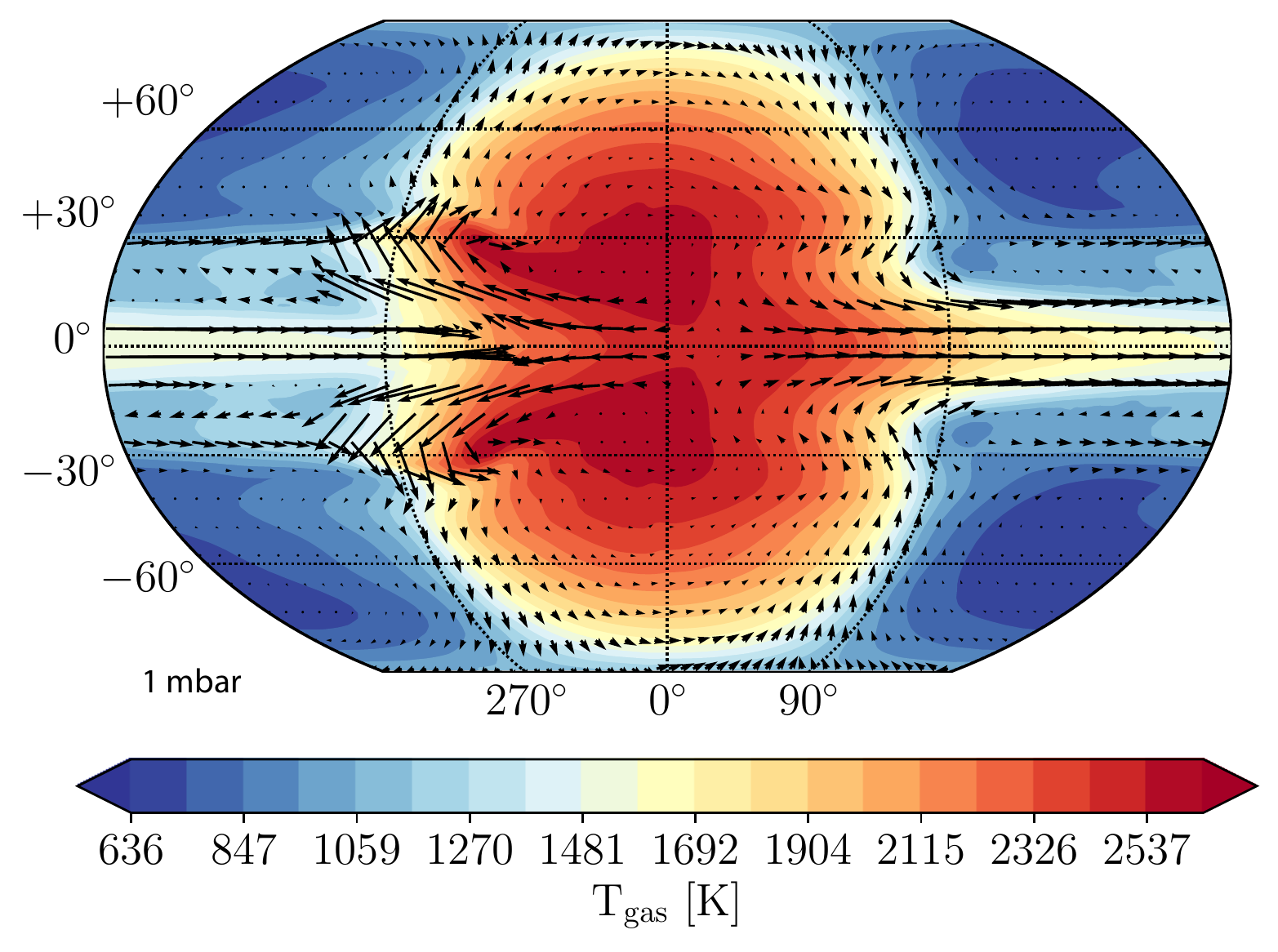}
   \includegraphics[width=0.49\textwidth]{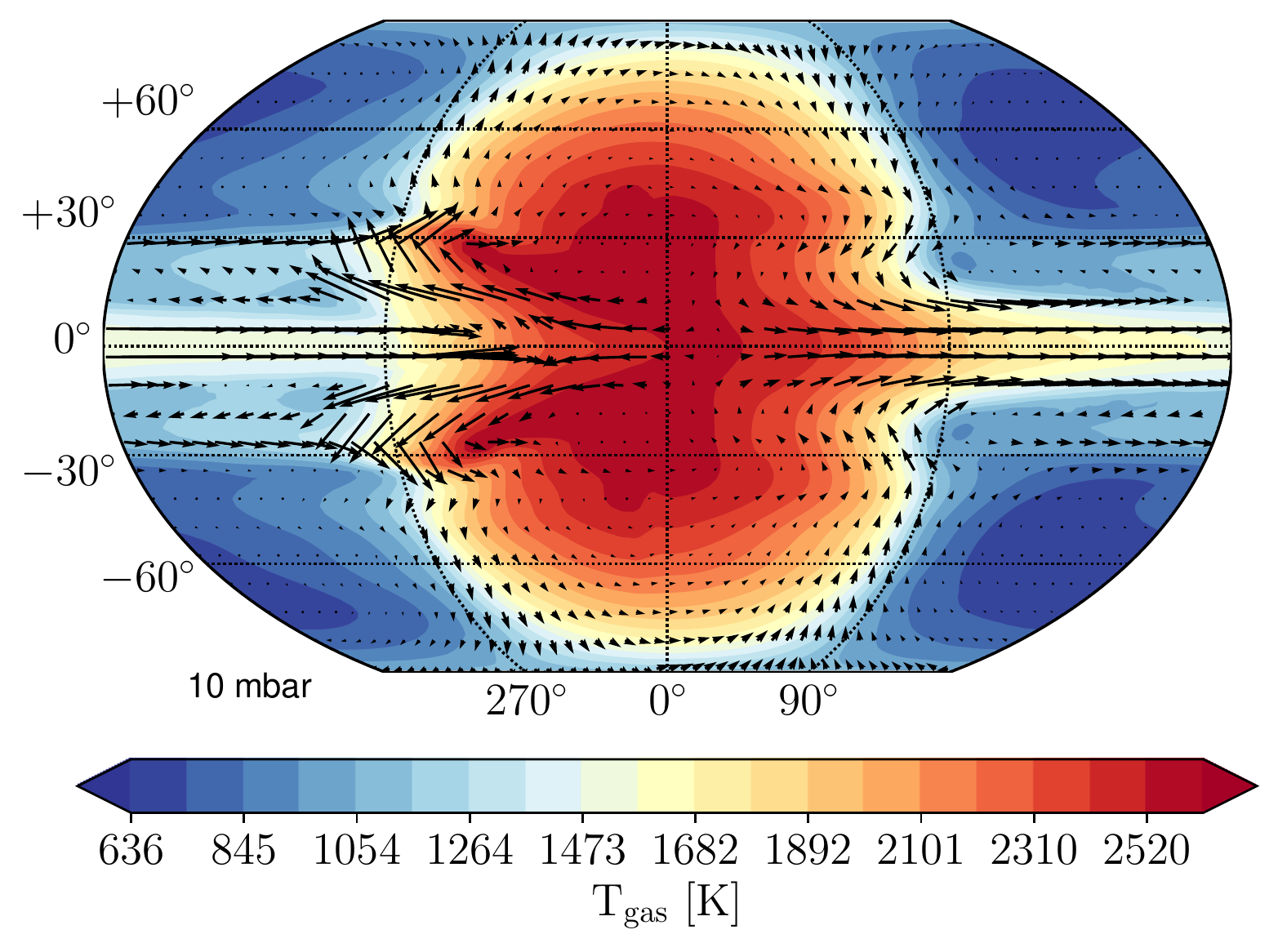}
   \includegraphics[width=0.49\textwidth]{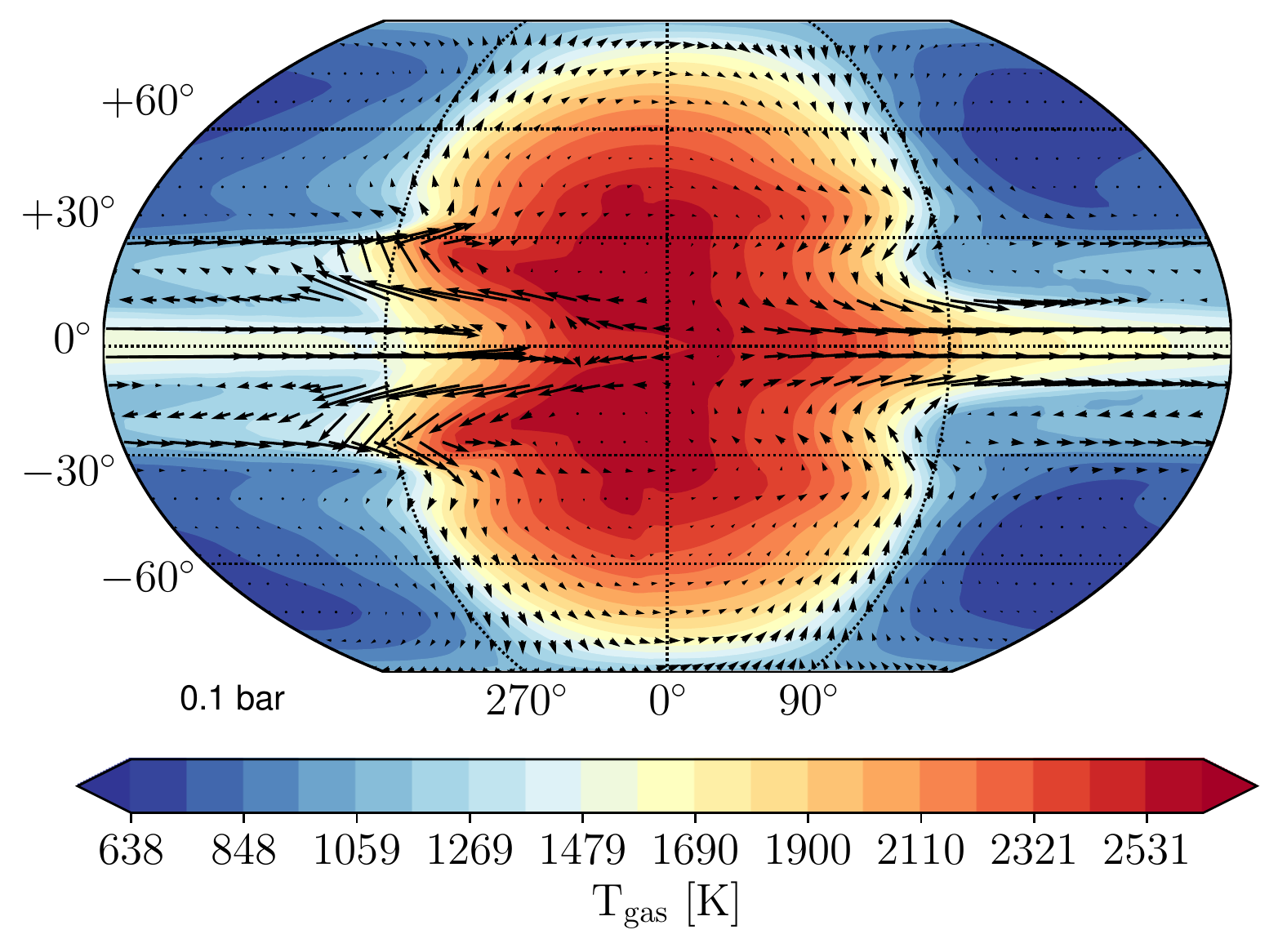}
   \includegraphics[width=0.49\textwidth]{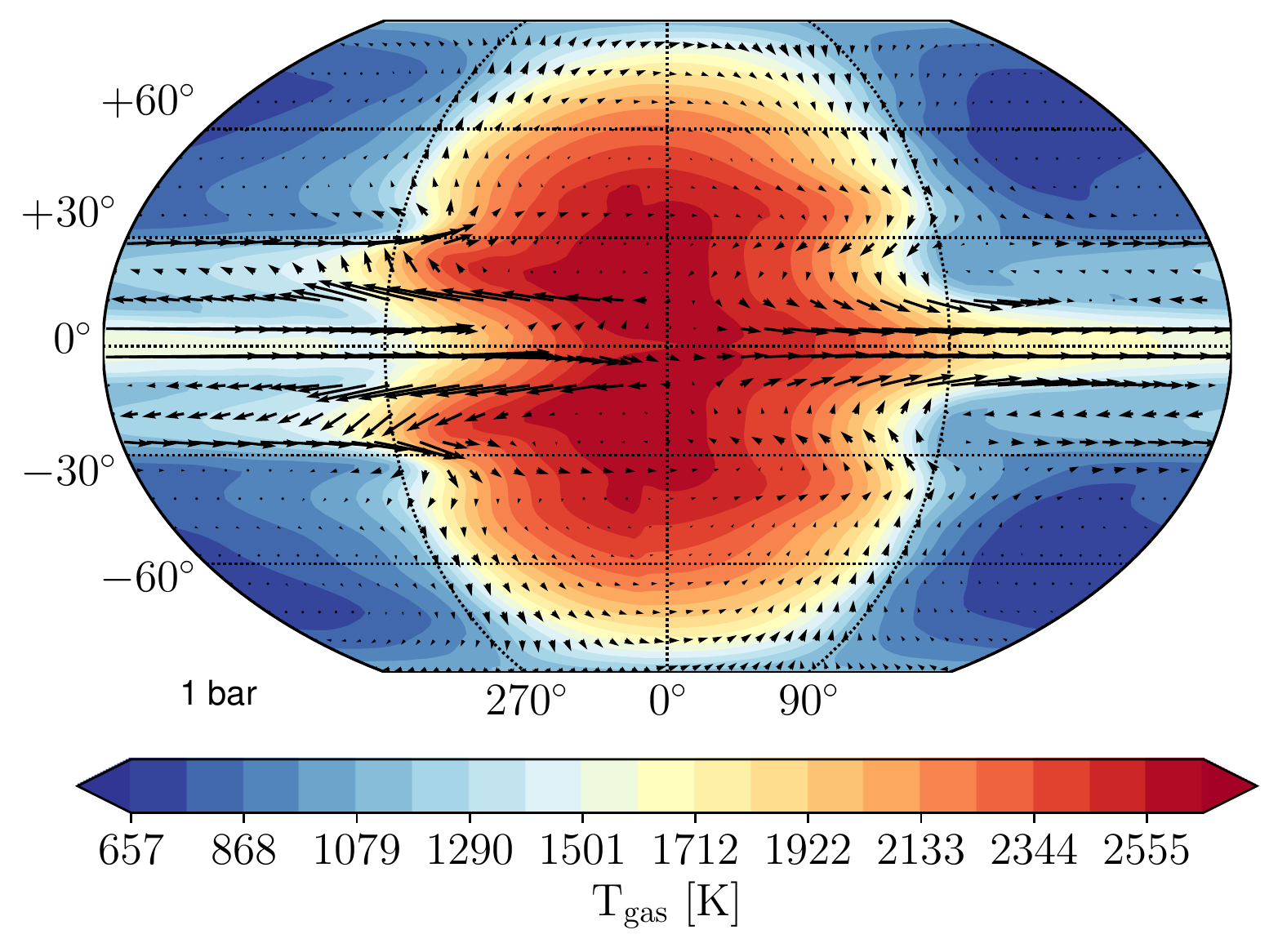}
   \includegraphics[width=0.49\textwidth]{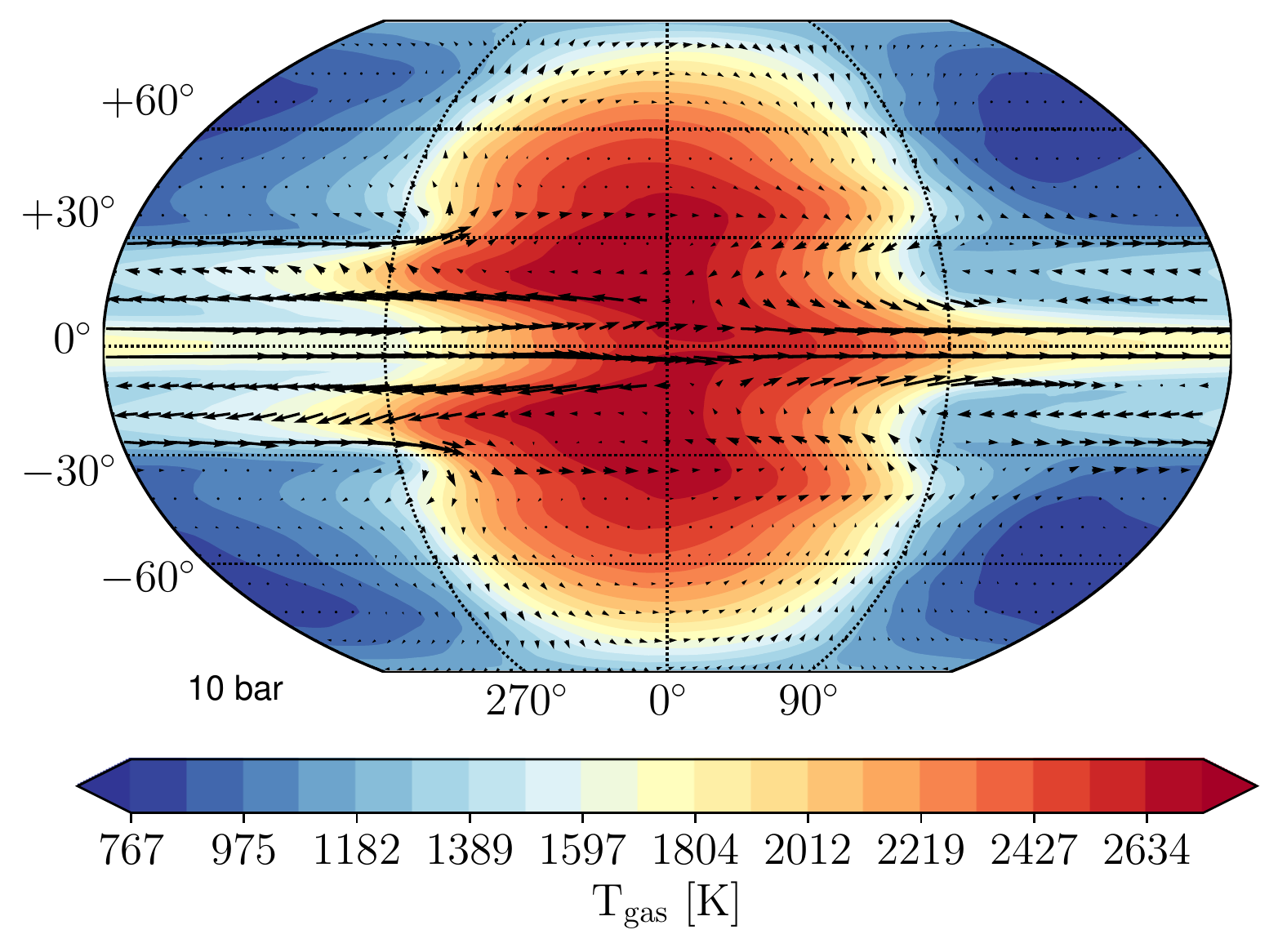}
   \includegraphics[width=0.49\textwidth]{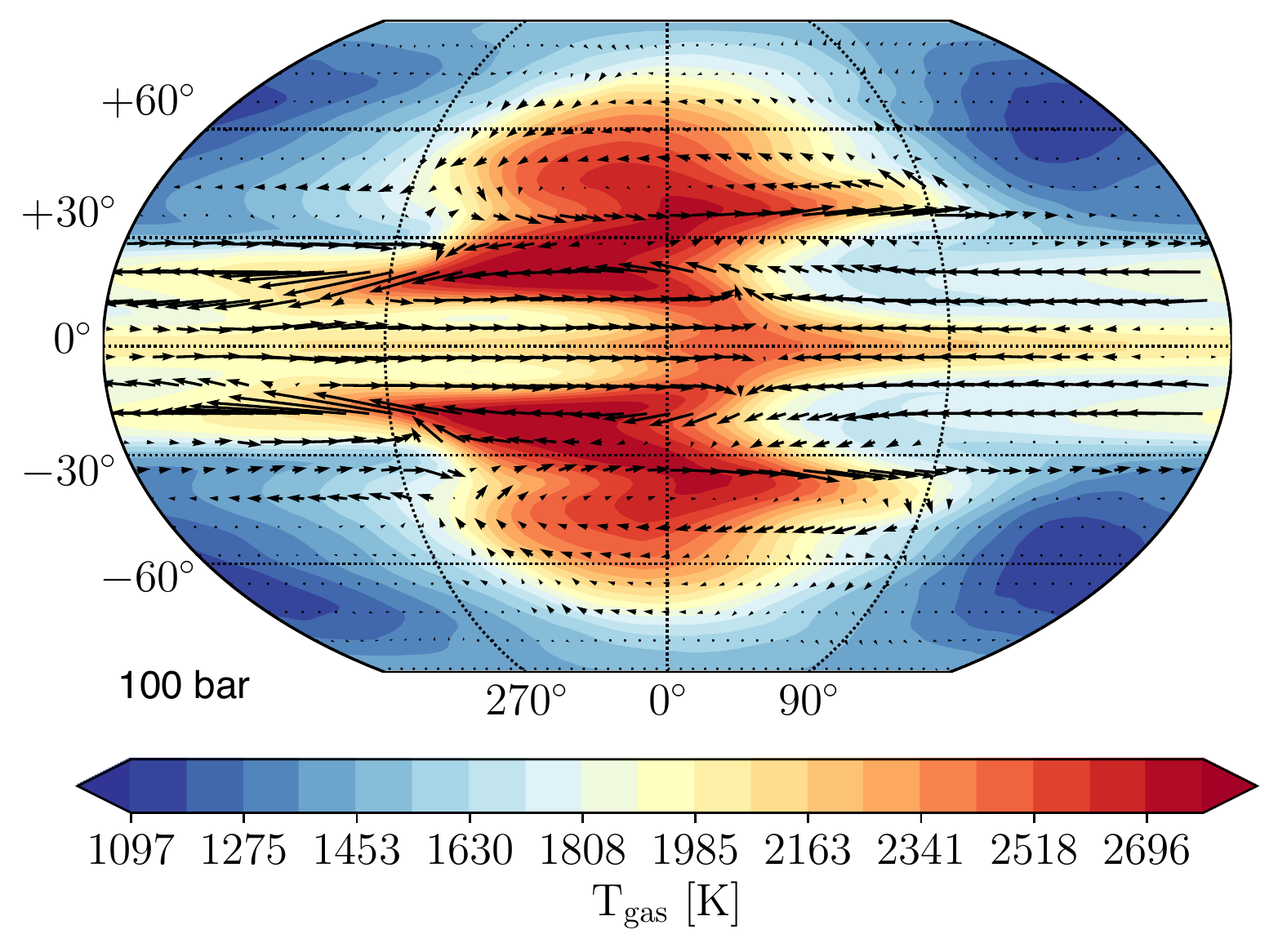}
   \caption{Atmospheric gas temperatures, T$_{\rm gas}$ [K], lat-lon maps at approximately 10$^{-3}$ and 0.01 bar (top row), 0.1 and 1 bar (middle row), 10 and 100 bar (bottom row) gas pressures.
   The vectors show the direction and relative magnitude of the wind speed.
   Note the scale is different between each plot.}
   \label{fig:map_T}
\end{figure*}

\begin{figure*} 
   \centering
   \includegraphics[width=0.49\textwidth]{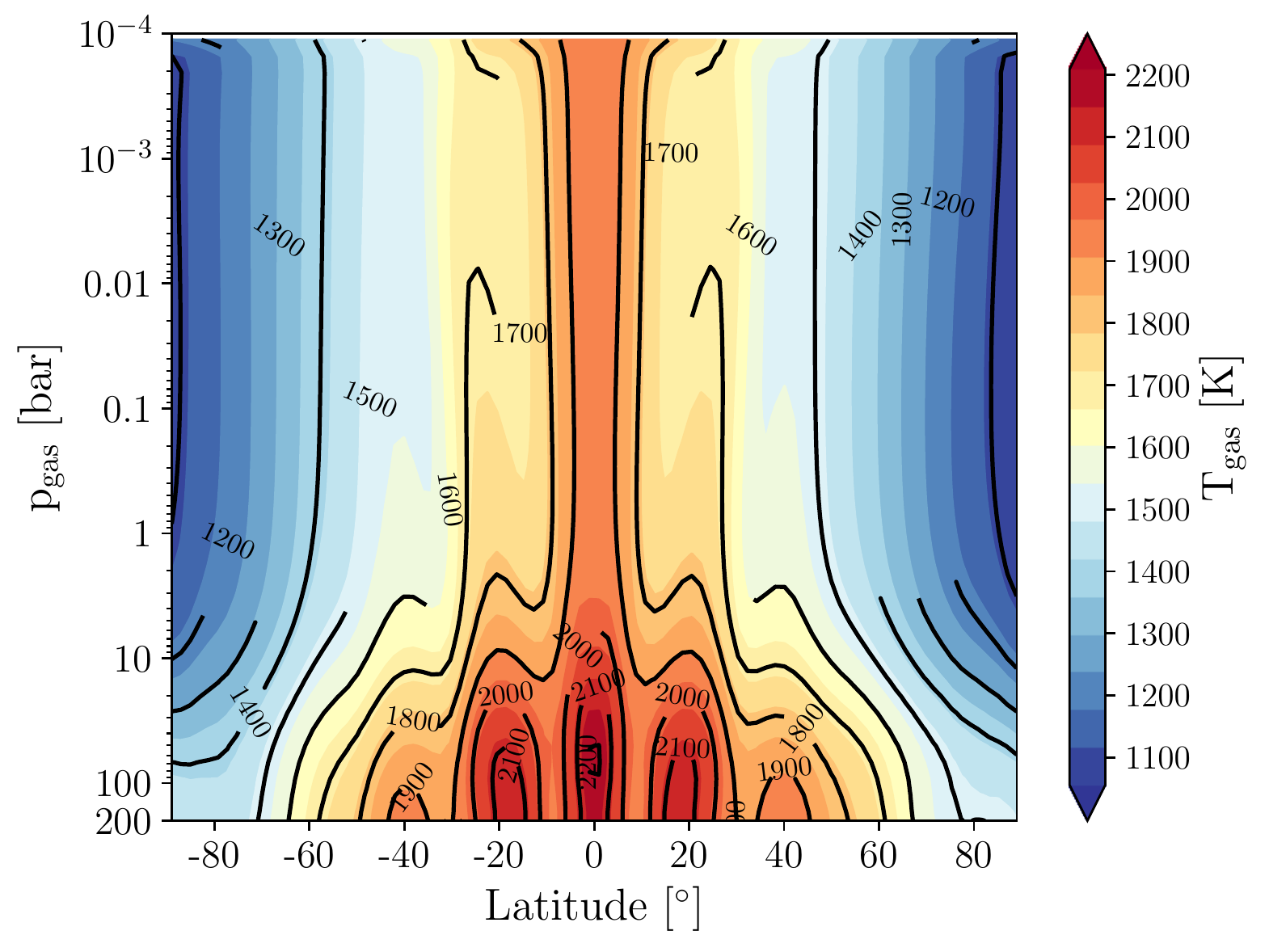}
   \includegraphics[width=0.49\textwidth]{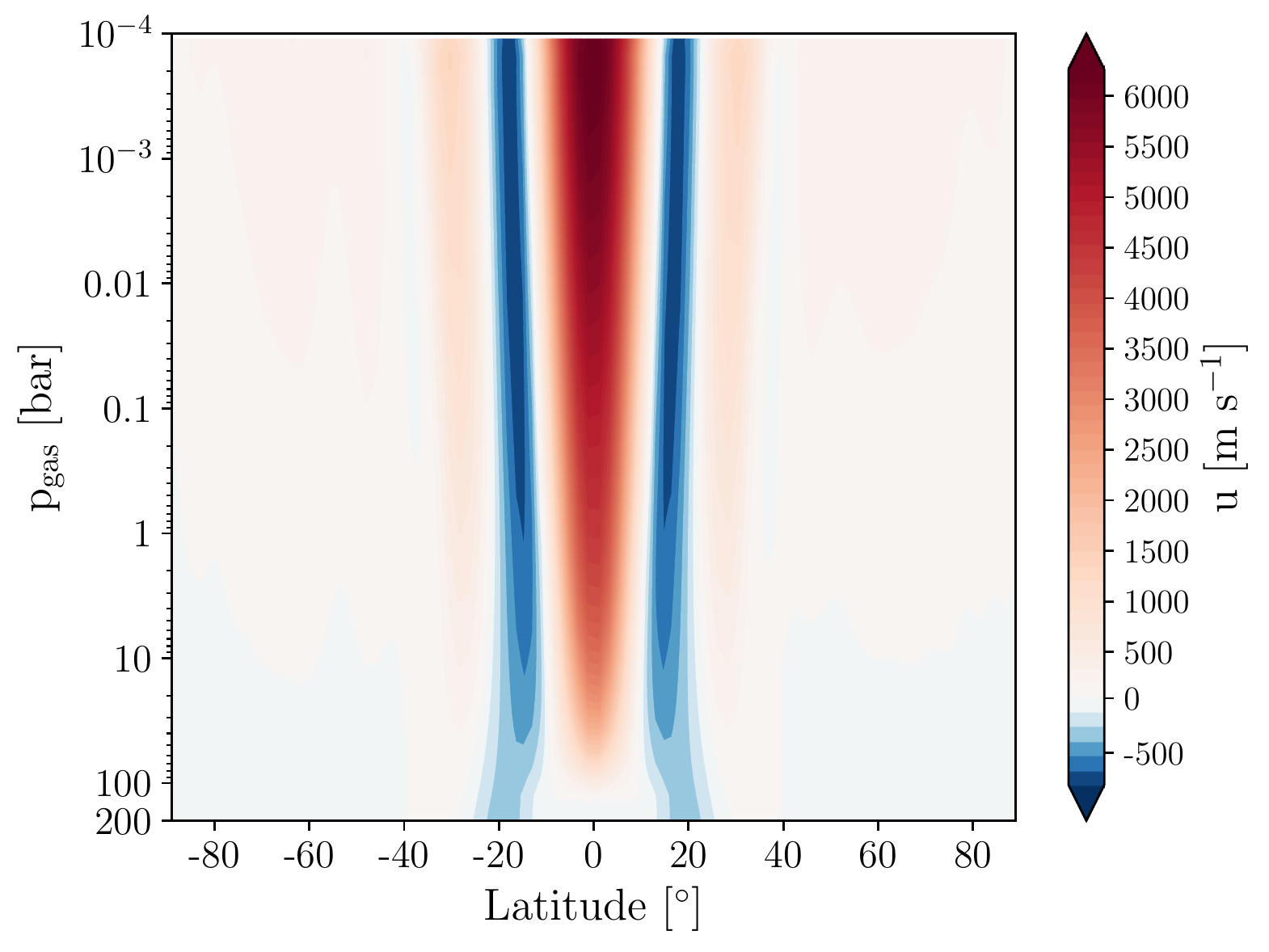}
   \includegraphics[width=0.49\textwidth]{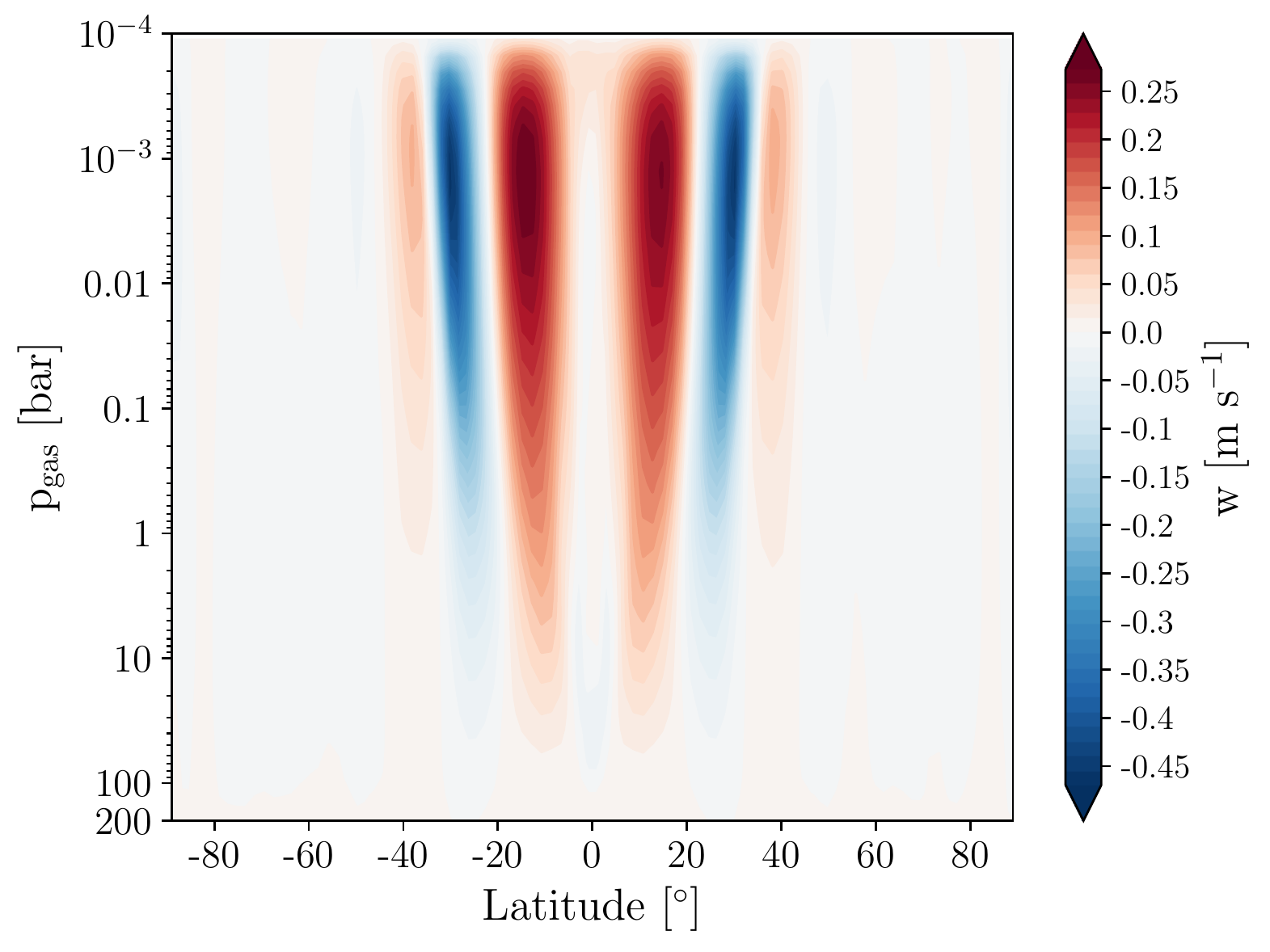}
   \includegraphics[width=0.49\textwidth]{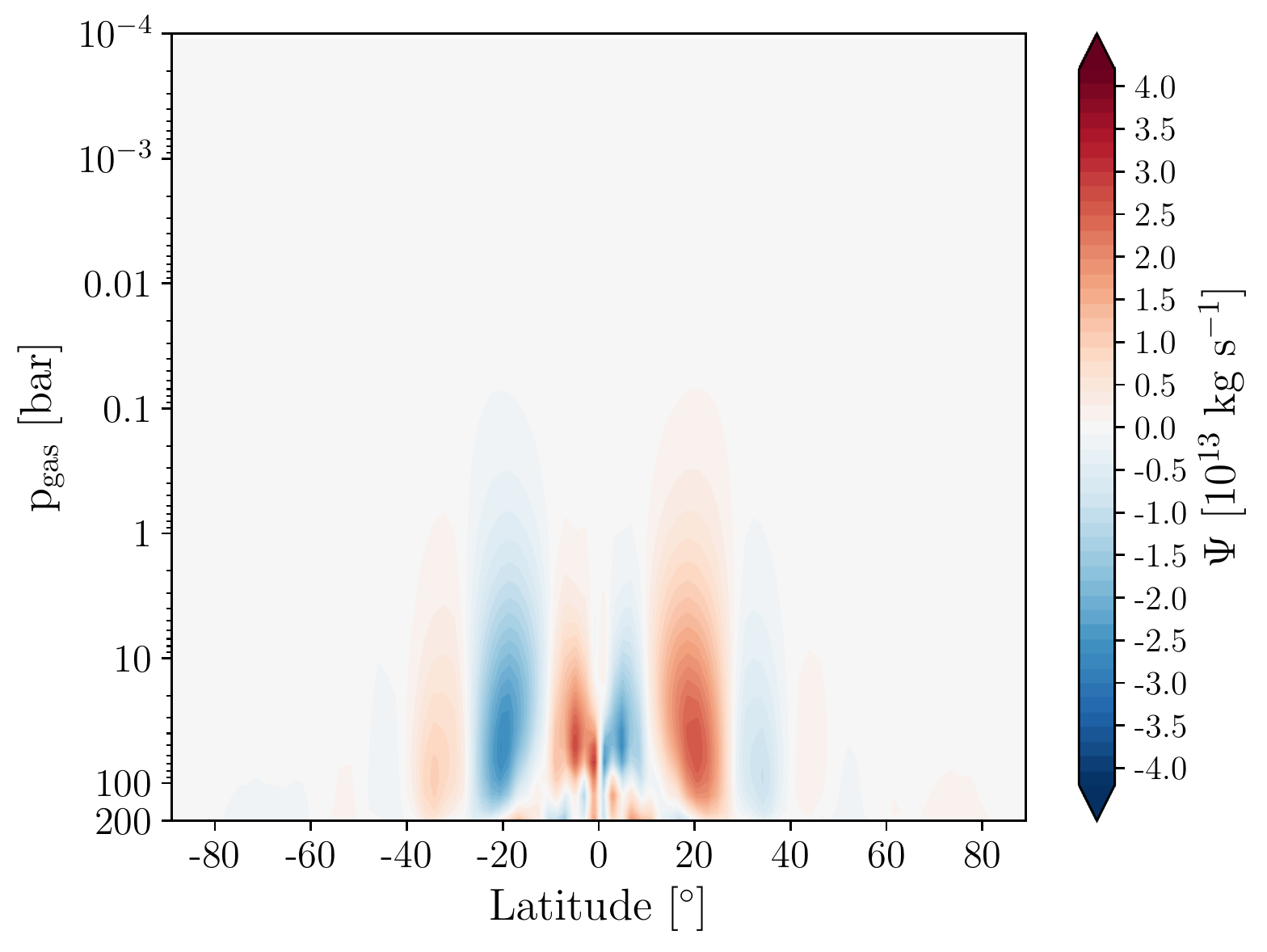}
   \caption{Zonal mean of the gas temperature (T$_{\rm gas}$ [K], top left), zonal velocity (u [m s$^{-1}$], top right), vertical velocity (w [m s$^{-1}$], lower left), and mass stream function ($\Psi$ [kg s$^{-1}$], lower right).}
   \label{fig:zmean_plots}
\end{figure*}

\begin{figure*} 
   \centering
   \includegraphics[width=0.49\textwidth]{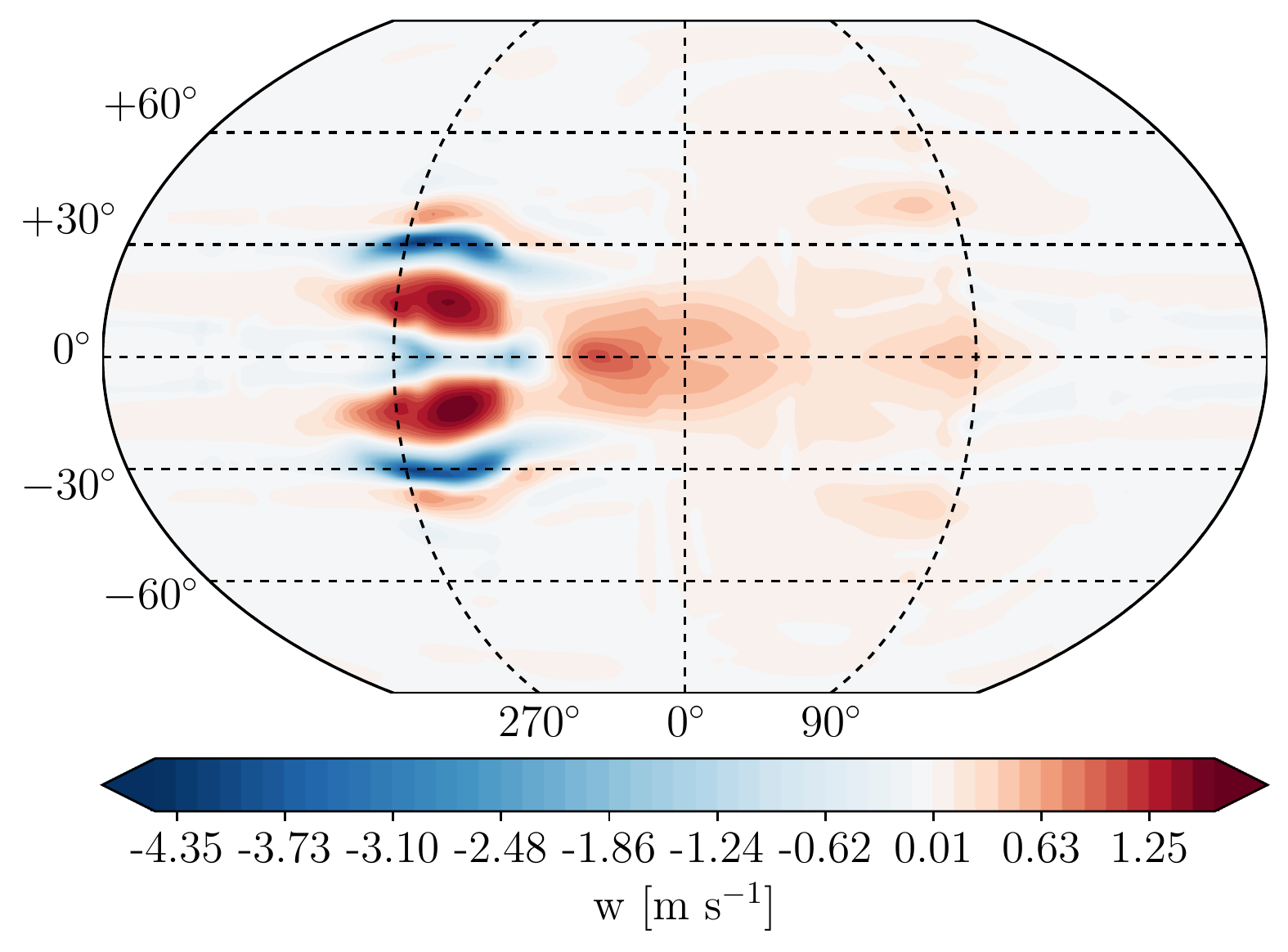}
   \includegraphics[width=0.49\textwidth]{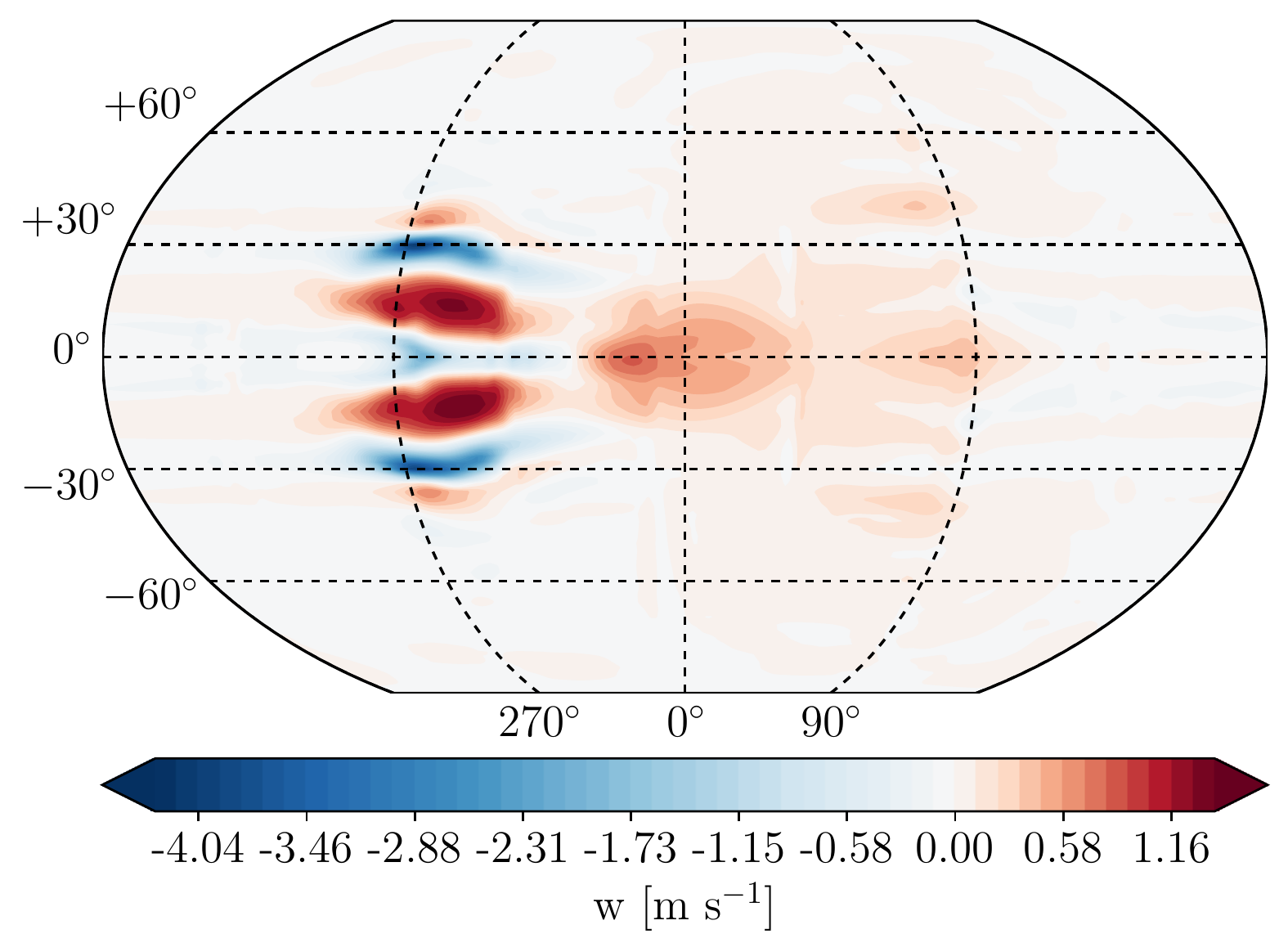}
   \includegraphics[width=0.49\textwidth]{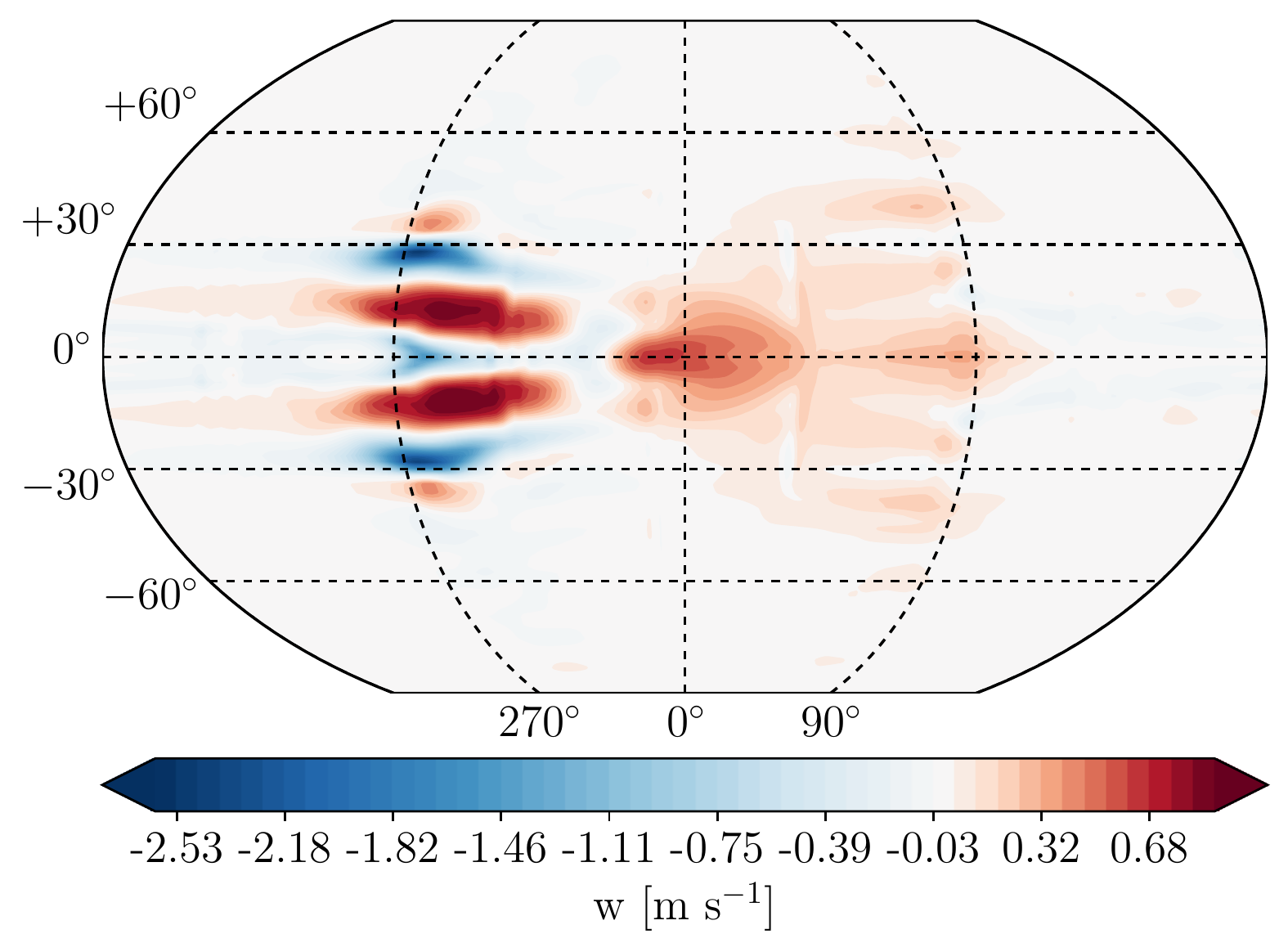}
   \includegraphics[width=0.49\textwidth]{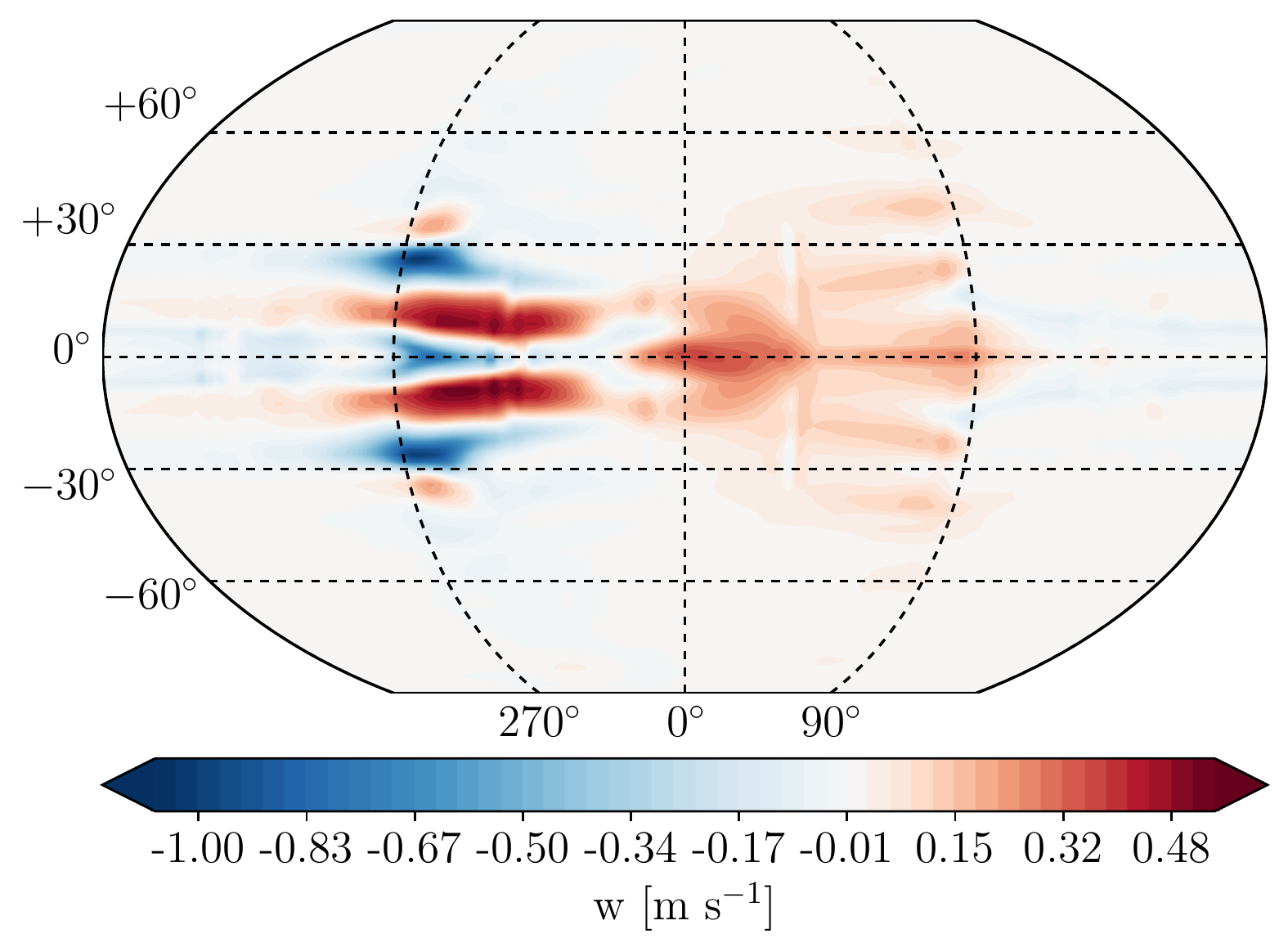}
   \includegraphics[width=0.49\textwidth]{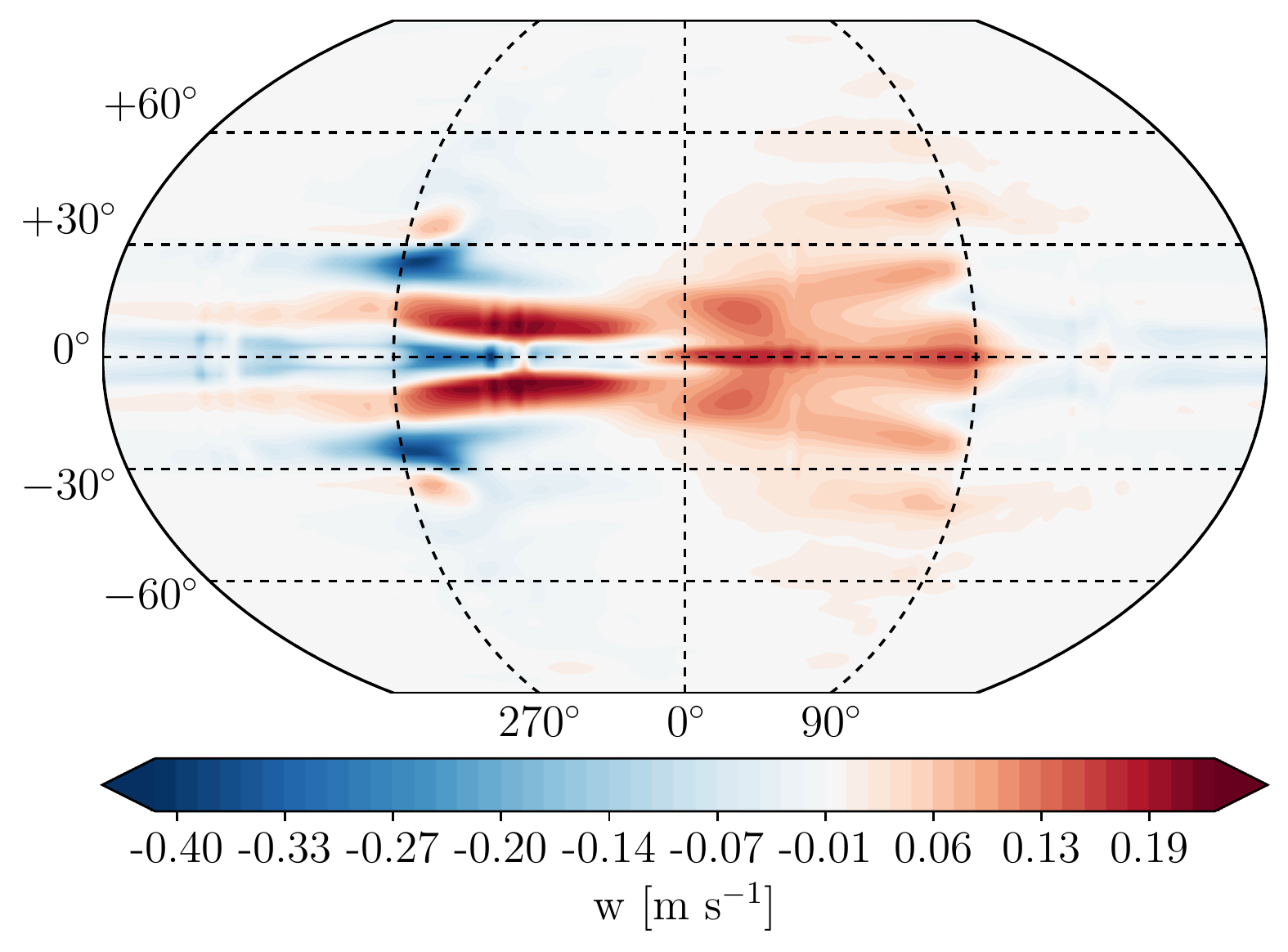}
   \includegraphics[width=0.49\textwidth]{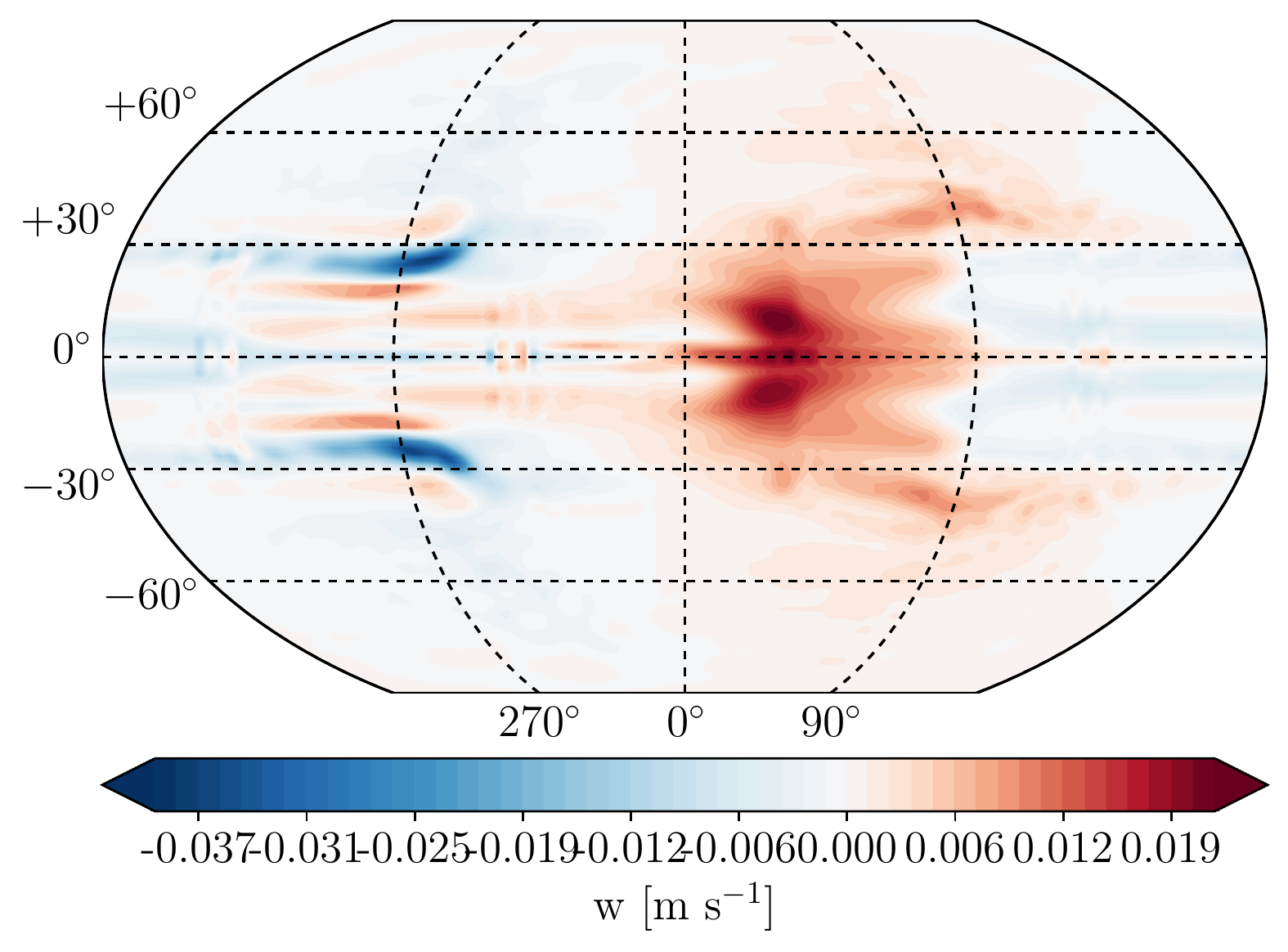}
   \caption{Atmospheric vertical velocity, w [m s$^{-1}$], lat-lon maps at approximately 10$^{-3}$ and 0.01 bar (top row), 0.1 and 1 bar (middle row), 10 and 100 bar (bottom row) gas pressures.
   Note the scale is different between each plot.}
   \label{fig:map_w}
\end{figure*}

In this section, we present the thermal and dynamical structures of the WD0137-349B GCM simulation.
Figure \ref{fig:Tp} shows the 1D temperature-pressure profiles at the equatorial region of the BD.
Dayside profiles are close to isothermal down to a pressure of $\approx$10 bar, where the atmosphere becomes optically thick in the optical band.
This suggests the atmosphere to be near radiative-equilibrium in most parts of the dayside atmosphere.

Figure \ref{fig:map_T} shows lat-lon maps of the temperature and velocity vector fields at pressure levels 10$^{-3}$, 0.01, 0.1, 1, 10 and 100 bar.
These maps show that the main redistribution of energy from the dayside to the nightside comes from the strong equatorial confined jet.
There is also significant westward shifting of hot spots from the dayside at latitudes of $\pm$ 20$^{\circ}$ from counter-rotating jets, and a slight shifting eastward at $\sim$ $\pm$ 30$^{\circ}$ latitude.

Figure \ref{fig:zmean_plots} presents the zonal mean temperature, zonal velocity, vertical velocity and mass stream function respectively.
The zonal mean temperature and zonal velocity plots show that efficient day-night energy transport is present at the equatorial regions of the model.
Regions at high latitudes, outside the main jet structures remain colder on average.
The zonal mean velocity plot suggests that global scale jets and counter-rotating jets are present.
However, the velocity vectors in the temperature map plots (Fig. \ref{fig:map_T}) suggest a more complex atmospheric wave structure, with a mixture of flows going with and against the rotation.
We briefly discuss these dynamical features in Sect. \ref{sec:discussion}.

The zonal mean vertical velocity plot suggests global scale upwelling and downwelling occurring inside the BD atmosphere.
However, in Fig. \ref{fig:map_w} we show the lat-lon pressure level maps of the vertical velocity.
These plots show that the downwelling is localised near the 270$^{\circ}$ longitude terminator and at the equator and $\pm$ 20$^{\circ}$ latitudes, while the majority of the upwelling is located on the dayside of the BD.
The zonal mass stream function plot along with the vertical velocity plots in Fig. \ref{fig:map_w} suggest multiple overturning structures at the 270$^{\circ}$ longitude terminator.

\subsection{OLR and atmospheric variability}
\label{sec:OLR}

\begin{figure*} 
   \centering
   \includegraphics[width=0.49\textwidth]{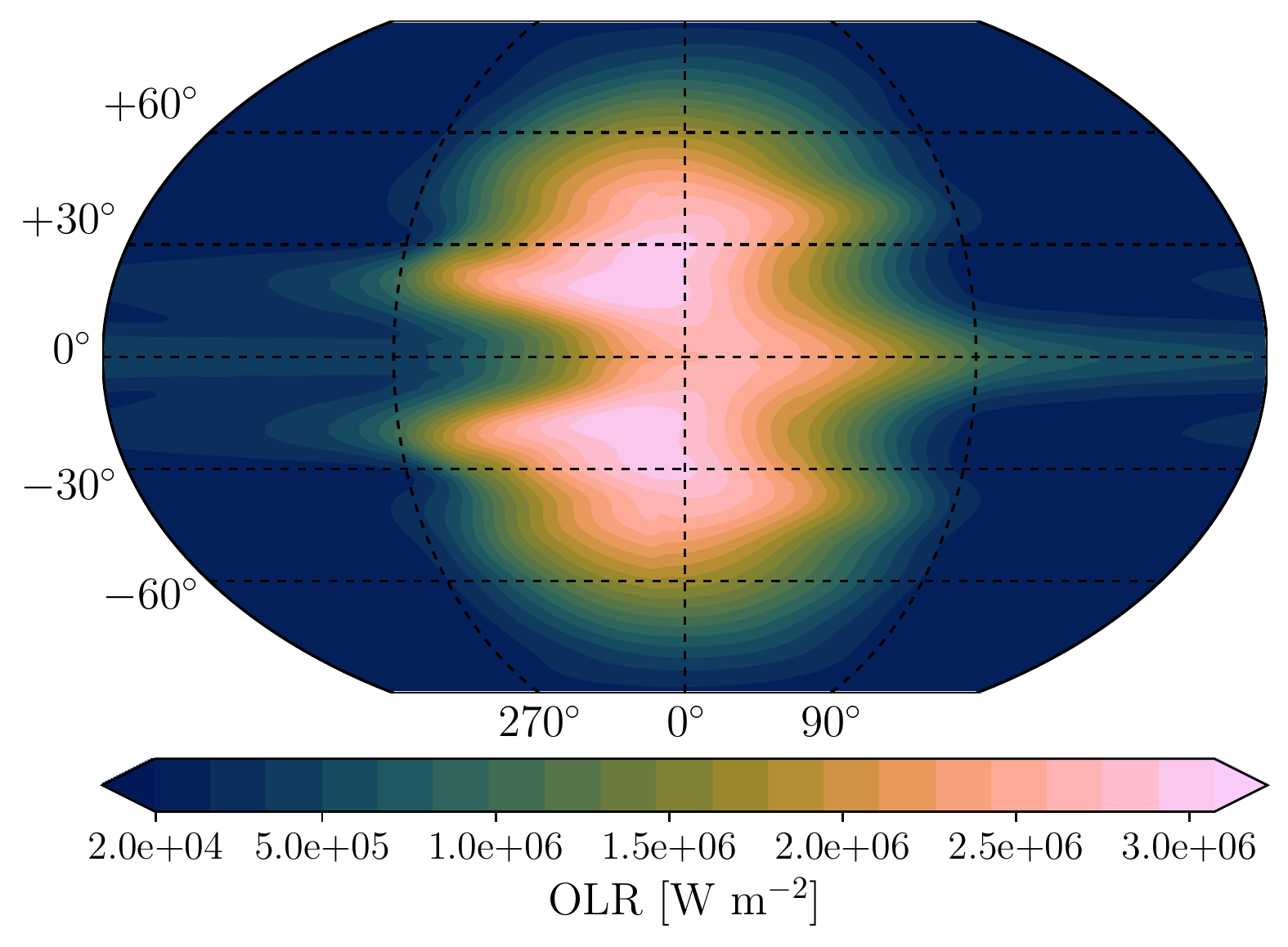}
   \includegraphics[width=0.49\textwidth]{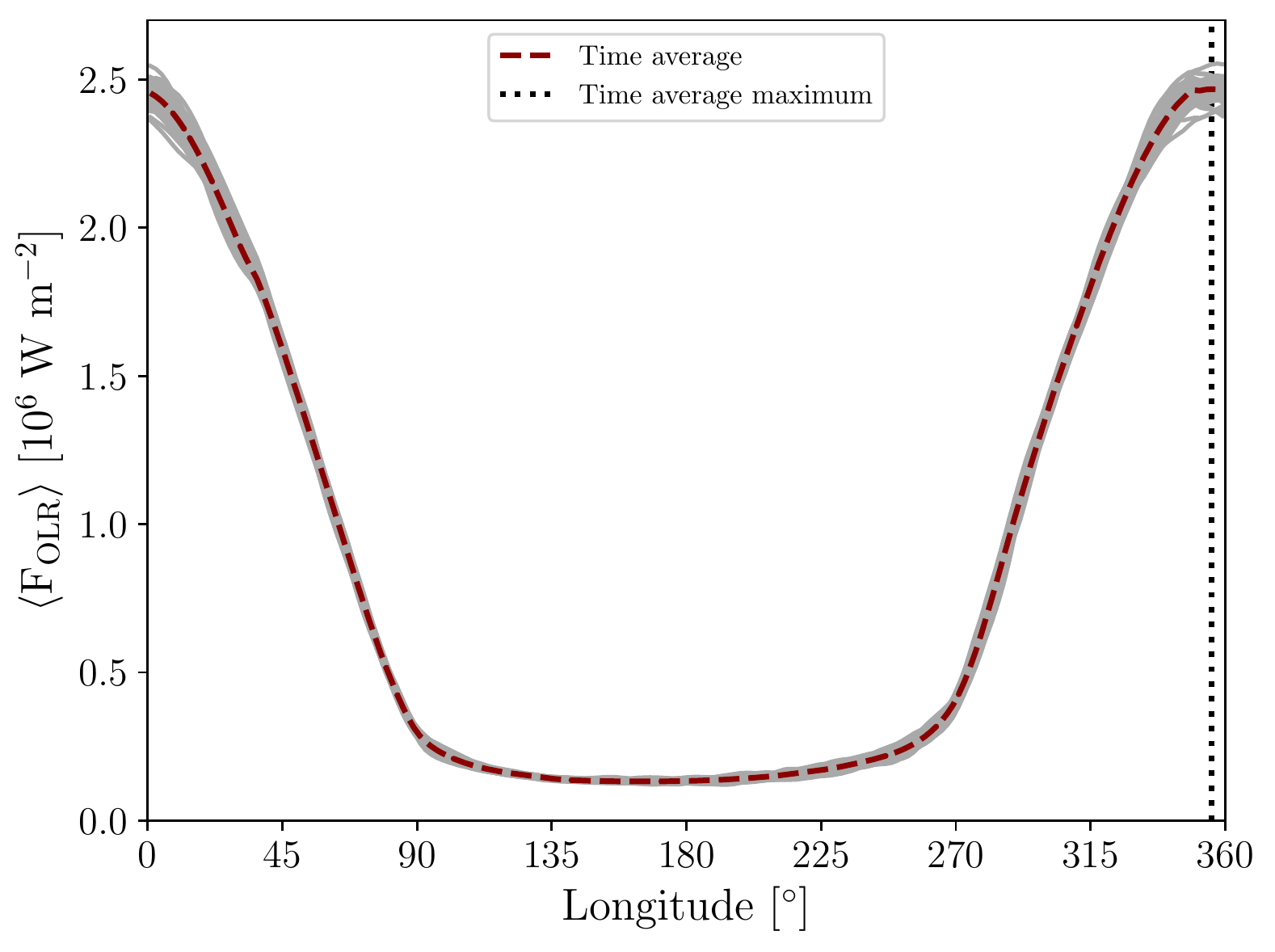}
   \caption{Left: Map of outgoing longwave radiation (OLR) flux [W m$^{-2}$] from the GCM model.
   Right: Latitudinally averaged OLR flux, $\langle$ F$_{\rm OLR}$$\rangle$ (Eq. \ref{eq:avF}), as a function of longitude
   The red dashed line shows the time averaged value, the grey lines show output every 10 days for the final 500 days of simulation and the vertical dotted line shows the longitude of the time averaged maximum flux.}
   \label{fig:olr}
\end{figure*}

Figure \ref{fig:olr} (left) presents the columnwise top of atmosphere (TOA) outgoing longwave radiative flux (OLR) of the averaged output.
The pattern corresponds well to the temperature structure from 1-10 bar (Fig \ref{fig:map_T}), the expected pressure levels where the longwave radiation becomes optically thin.

To examine the variability in our model, we calculate the latitudinally averaged OLR flux, $\langle$F$_{\rm OLR}$$\rangle$ [W m$^{-2}$], given by \citep[e.g.][]{Heng2011b}
\begin{equation}
\label{eq:avF}
\langle F_{\rm OLR}\rangle = \frac{1}{\pi}\int_{-\pi/2}^{\pi/2}F_{\rm OLR} \cos^{2}\Phi \ \textrm{d}\Phi,
\end{equation}
where F$_{\rm OLR}$ [W m$^{-2}$] is the columnwise OLR flux from the GCM model.
On the RHS of Fig. \ref{fig:olr} we present the averaged flux from Eq. \ref{eq:avF} as a function of longitude for every 10 days for the final 500 days of simulation.
The OLR variation is $\lesssim$0.25 $\times$ 10$^{6}$ W m$^{-2}$ on the dayside phases of the BD, while nightside phases remain relatively constant with time.
A slight westward shift ($\approx$ 5$^{\circ}$) in the maximum OLR is also present as shown by the vertical dotted line.

\section{Post-processing and comparison to observations}
\label{sec:post-processing}

In this section, we produce synthetic emission spectra and phase curves from the GCM results to compare directly to the available observational data.
We apply the hybrid raytracing and 3D Monte Carlo radiative-transfer model \textsc{cmcrt} \citep{Lee2017} in correlated-k mode \citep{Lee2019} to calculate the output emission spectra of the GCM.
Due to the strong day-night temperature contrast, we apply the composite emission biasing of \citet{Baes2016} with a $\xi_{\rm em}$ bias coefficient of 0.99.
We also develop a biasing scheme for sampling the k-coefficients in emission, detailed in App. \ref{app:1}, based on the \citet{Baes2016} methodology.
To avoid spurious noise from the inverted temperature profiles near the uppermost boundary layers (e.g. Fig. \ref{fig:Tp}), the temperature of the top two layers are assumed to be equal to the third most upper layer.

The volume mixing ratio of molecular and elemental species is calculated assuming chemical equilibrium (CE) using the \textsc{GGChem} code \citep{Woitke2018} at the solar elemental ratios from \citet{Asplund2009}.
We include the calculation of thermally ionised species in the CE calculation to more accurately capture the chemical structure of the hotter (T$_{\rm gas}$ $>$ 2000 K) atmospheric regions.

A key difference between typical HJ phase curve modelling and this system is the low inclination of WD0137-349B ($\sim$35$^{\circ}$), resulting in different higher latitude fractions of the BD dayside and nightside regions in the observational line of sight at each phase.
\textsc{cmcrt} takes this into account by calculating the viewing angles as a function of phase for a 35$^{\circ}$ system inclination (i.e. viewing the planet at latitude of +55$^{\circ}$).
To produce combined WD + BD fluxes we use the same WD model as in \citet{Casewell2015}, originally produced from the \textsc{tlusty} and \textsc{synspec} models \citep{Hubeny1988,Hubeny1995}.
The WD fluxes were convolved with the \textit{H}, \textit{J}, \textit{K$_{s}$} and \textit{Spitzer} filter profiles to calculate the WD flux in each band.

\subsection{Input opacities}

Our molecular k-coefficients are calculated from the ExoMol database \citep{Tennyson2016} line lists with H$_{2}$ pressure broadening at a resolution of R1000 between 0.3 and 300 $\mu$m.
For Na and K we take the line list from the NIST database \citep{NISTWebsite} and broadening profile based on \citet{Allard2007b}.
Table \ref{tab:line-lists} contains all opacity sources used in the \textsc{cmcrt} simulation and their associated references.
The opacity of each GCM cell is calculated by interpolating from the table of k-coefficients of each species and combined using the random overlap method \citep[e.g.][]{Lacis1991, Amundsen2017}.

\begin{table}
\centering
\caption{Opacity sources and references used in the \textsc{cmcrt} post-processing of the GCM model output.}
\begin{tabular}{c c}  \hline \hline
Opacity Source & Reference   \\ \hline
Line & \\ \hline
Na   &  \citet{NISTWebsite}   \\
K   &  \citet{NISTWebsite}  \\
H$_{2}$O   &  \citet{Polyansky2018}   \\
CH$_{4}$   &  \citet{Yurchenko2017}  \\
NH$_{3}$   &  \citet{Yurchenko2011}   \\
CO   &  \citet{Li2015}   \\
CO$_{2}$   &  \citet{Rothman2010}   \\ \hline
Collision induced absorption & \\ \hline
H$_{2}$-H$_{2}$ & \citet{Baudino2017} \\
H$_{2}$-He & \citet{Baudino2017} \\ \hline
Rayleigh scattering & \\ \hline
H$_{2}$ & \citet{Irwin2009} \\
He  & \citet{Irwin2009} \\
\hline
\end{tabular}
\label{tab:line-lists}
\end{table}

\subsection{Emission spectra}

\begin{figure*} 
   \centering
   \includegraphics[width=0.49\textwidth]{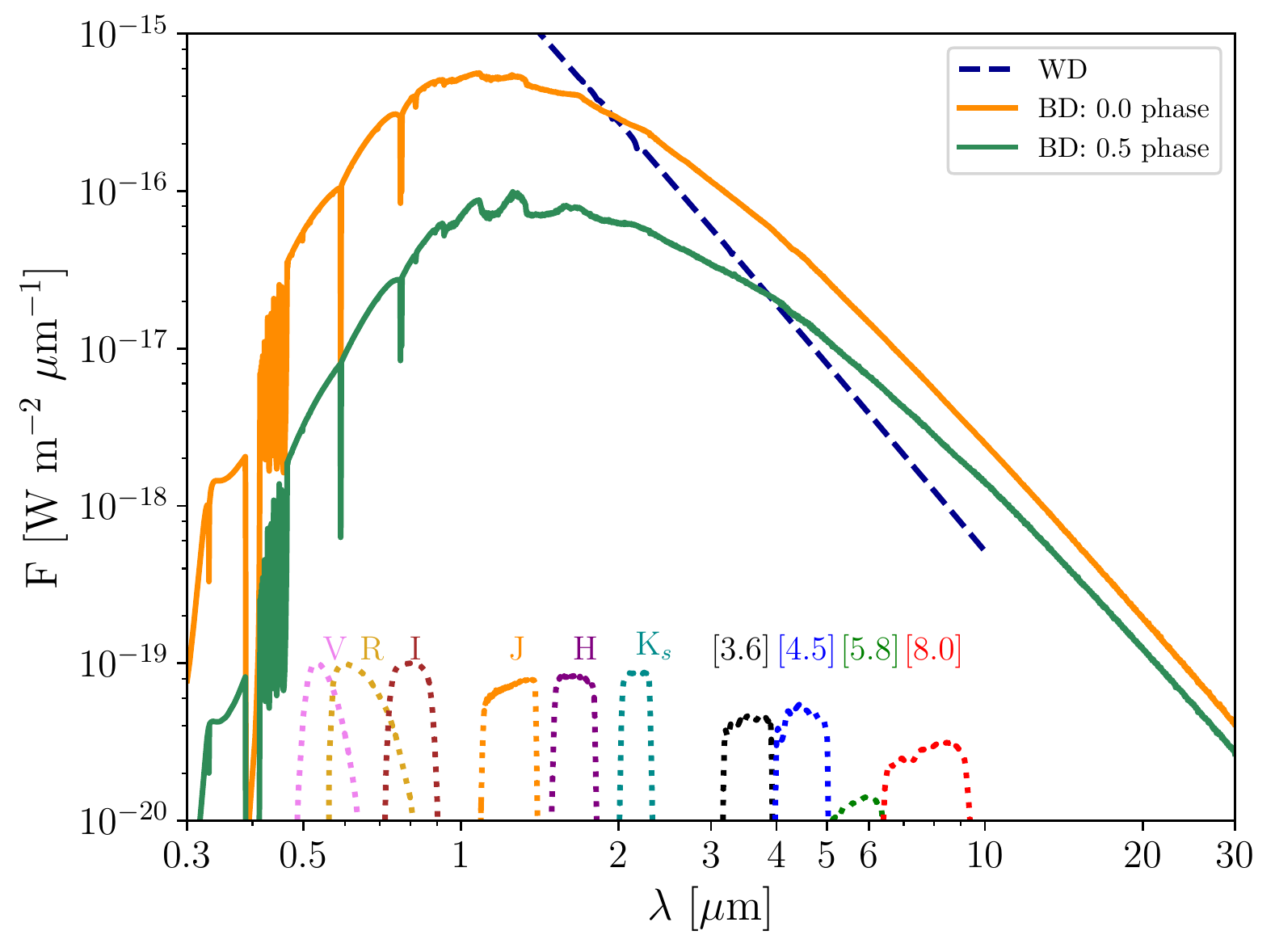}
   \includegraphics[width=0.49\textwidth]{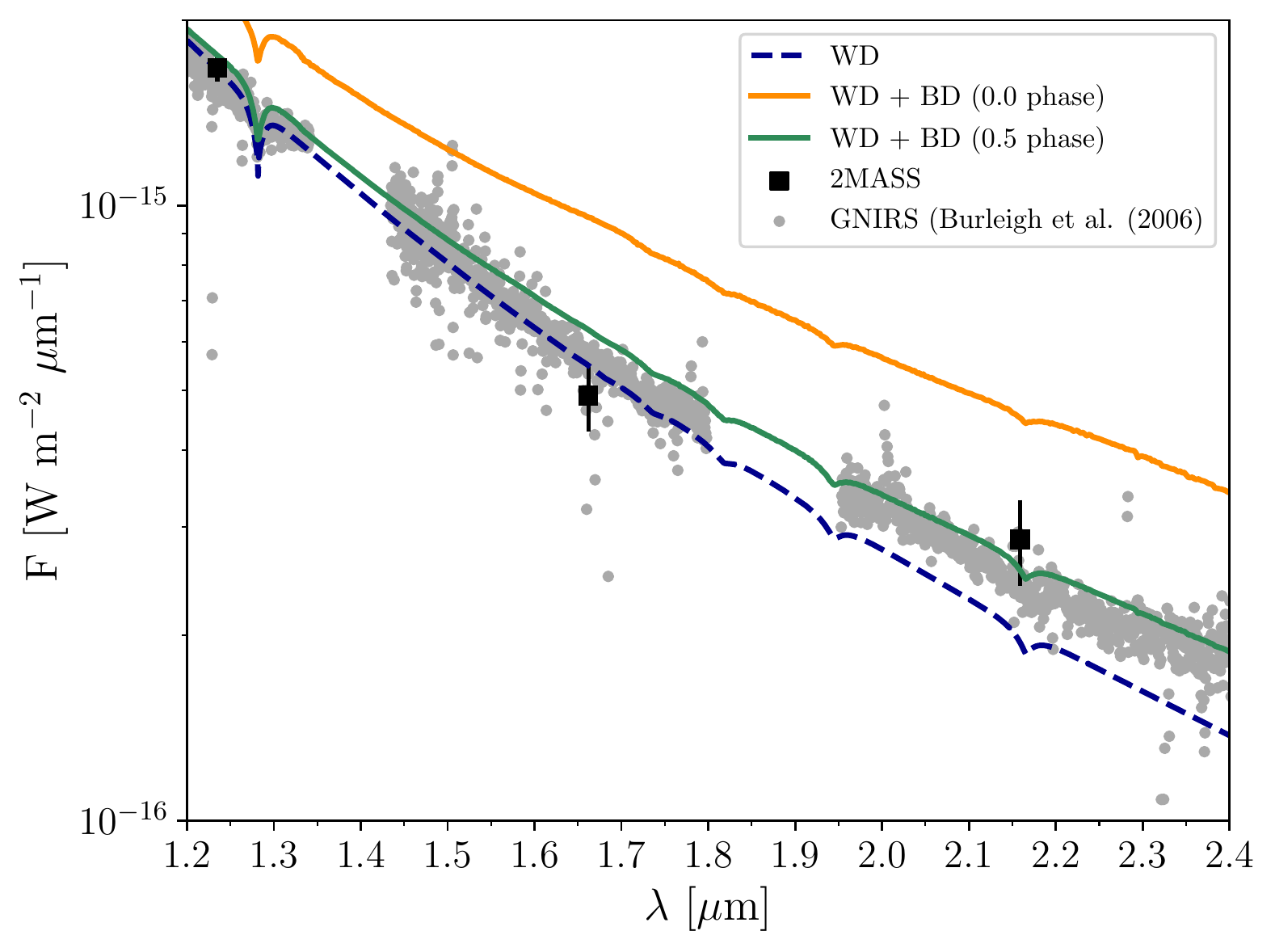}
   \caption{Left: post-processed emission spectra of the BD model at 0 (orange) and 0.5 (green) phases.
   Right: model emission spectra between 1.2 and 2.4 $\mu$m with archival 2MASS photometry data (black points) and the GNIRS spectra from \citet{Burleigh2006} (grey points).
   The modelled WD flux is the dashed blue line.
   The bandpasses used in \citet{Casewell2015} are plotted as dotted lines near the x-axis.}
   \label{fig:em_spec_2}
\end{figure*}

Figure \ref{fig:em_spec_2} (LHS) shows the emission spectrum of the model WD and the post-processed GCM output at 0 and 0.5 phase.
The observational bandpasses of the instruments used in \citet{Casewell2015} are also plotted.
Each band is sensitive to the opacity features of different molecules considered in this study;
\begin{itemize}
\item \textit{V} band: Na
\item \textit{R} band: Na \& K
\item \textit{I} band: K
\item \textit{J} band: H$_{2}$O, CH$_{4}$, NH$_{3}$
\item \textit{H} band: H$_{2}$O, NH$_{3}$
\item \textit{K$_{s}$} band: CH$_{4}$, NH$_{3}$
\item \textit{Spitzer} 3.6: CH$_{4}$
\item \textit{Spitzer} 4.5: CO, CO$_{2}$
\item \textit{Spitzer} 5.8: H$_{2}$O
\item \textit{Spitzer} 8.0: H$_{2}$O, CH$_{4}$, NH$_{3}$
\end{itemize}
Any absorption features in the model emission spectra or modulations seen in the phase curves are therefore a convolution of the differences in the photospheric temperatures as a function of phase and also any change in the chemical composition between hemispheres.

We reproduce the start of the near-IR excess at $\approx$1.95 $\mu$m observed at the nightside phase of the BD in \citet{Burleigh2006} (RHS Fig. \ref{fig:em_spec_2}), suggesting that the fraction of the dayside and nightside emission of the BD at an orbital inclination of 35$^{\circ}$ is a reasonable approximation, rather than the expected T$_{\rm eq}$ $\sim$ 2000 K emitted flux at a 90$^{\circ}$ inclination.

Our model is unable to reproduce the Na and K emission features reported on the dayside of the BD by \citet{Longstaff2017}.
This is due to a lack of an upper atmosphere temperature inversion present in the GCM thermal structures.
We suggest possible mechanisms to produce such an inversion in Sect. \ref{sec:discussion}.

\subsection{Phase curve comparisons}

\begin{figure*} 
   \centering
   \includegraphics[width=0.49\textwidth]{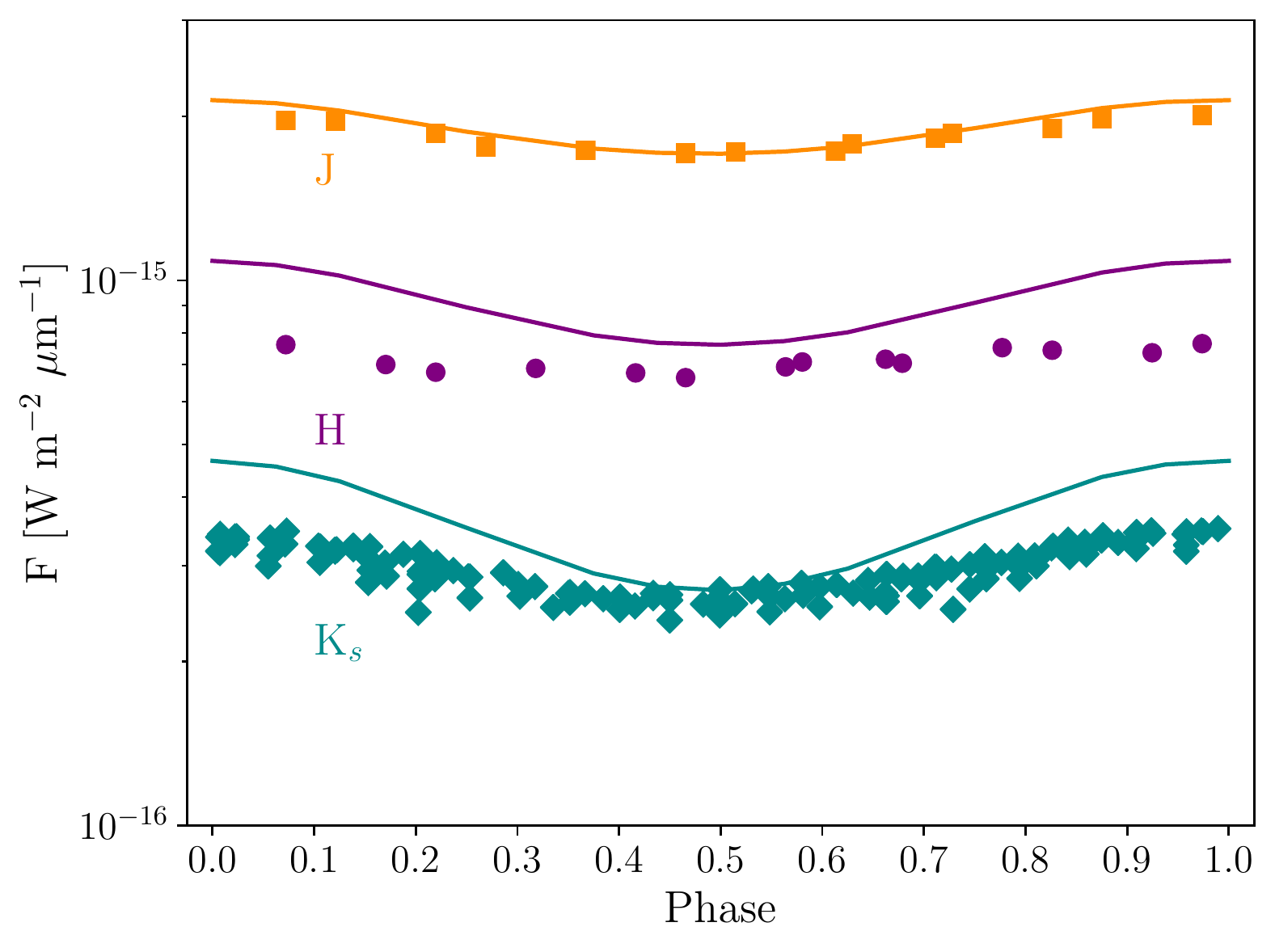}
   \includegraphics[width=0.49\textwidth]{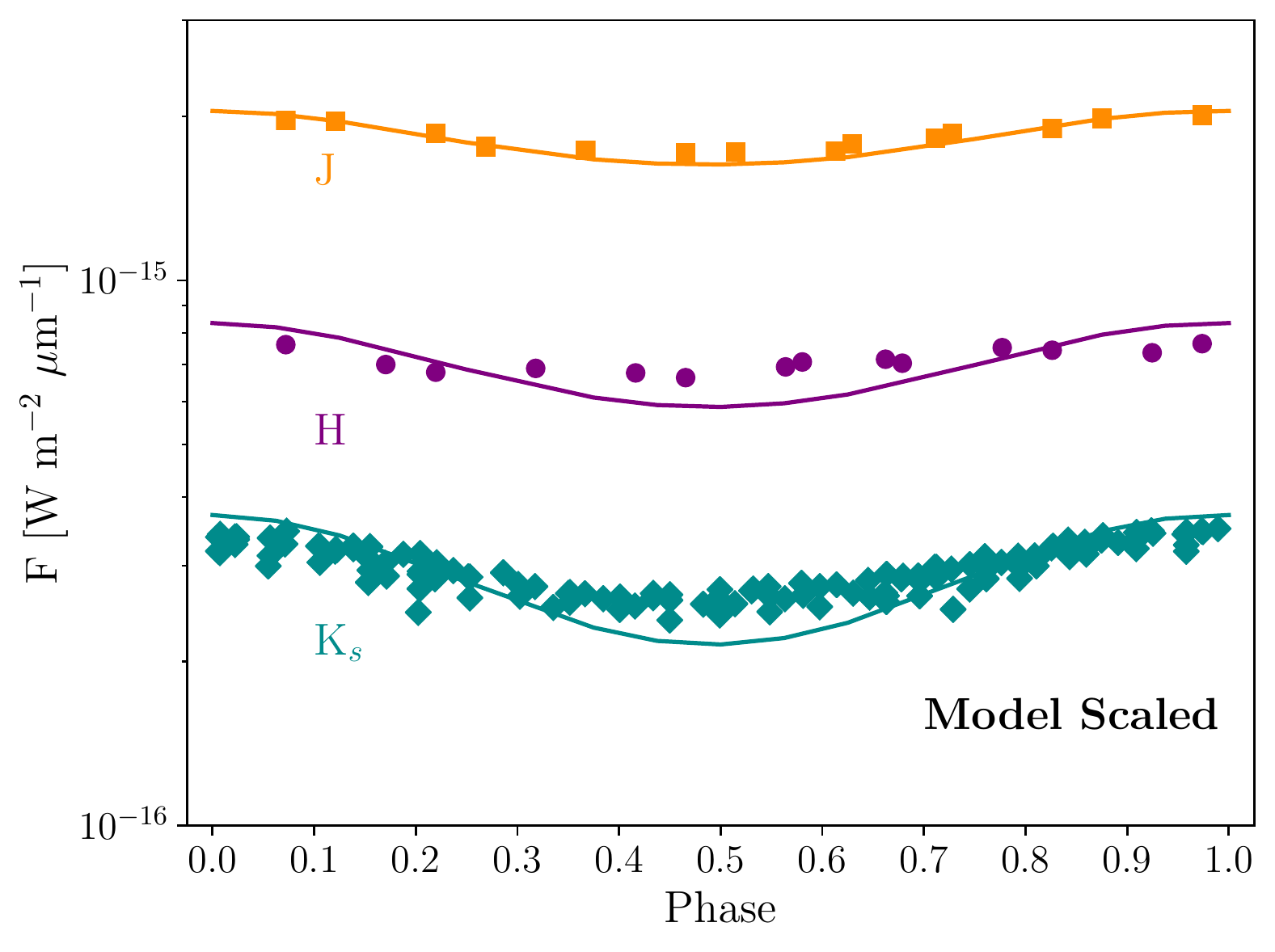}
   \caption{\textit{J} (orange), \textit{H} (purple) and \textit{K$_{s}$} (cyan) band phase curves from \citet{Casewell2015} (points) and post-processed GCM model fluxes (lines).
   The left plot shows the fiducial model fluxes while the right shows the model fluxes normalised to the average observed flux in each band, given by the relative factors in Table \ref{tab:shifts}.
   Errors in the observational data are on the order of the point size.}
   \label{fig:phase_JHK}
\end{figure*}

\begin{figure*} 
   \centering
   \includegraphics[width=0.49\textwidth]{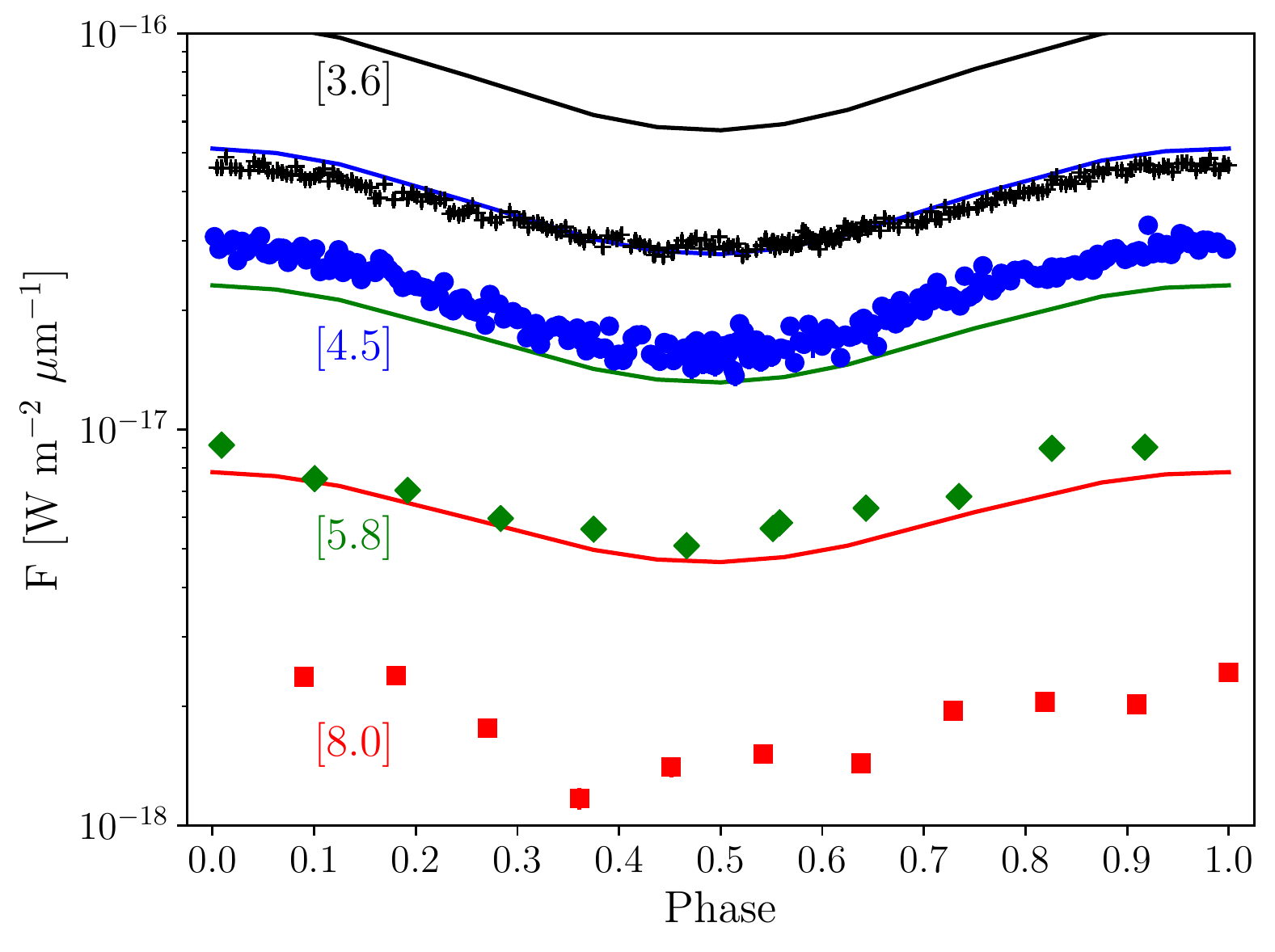}
   \includegraphics[width=0.49\textwidth]{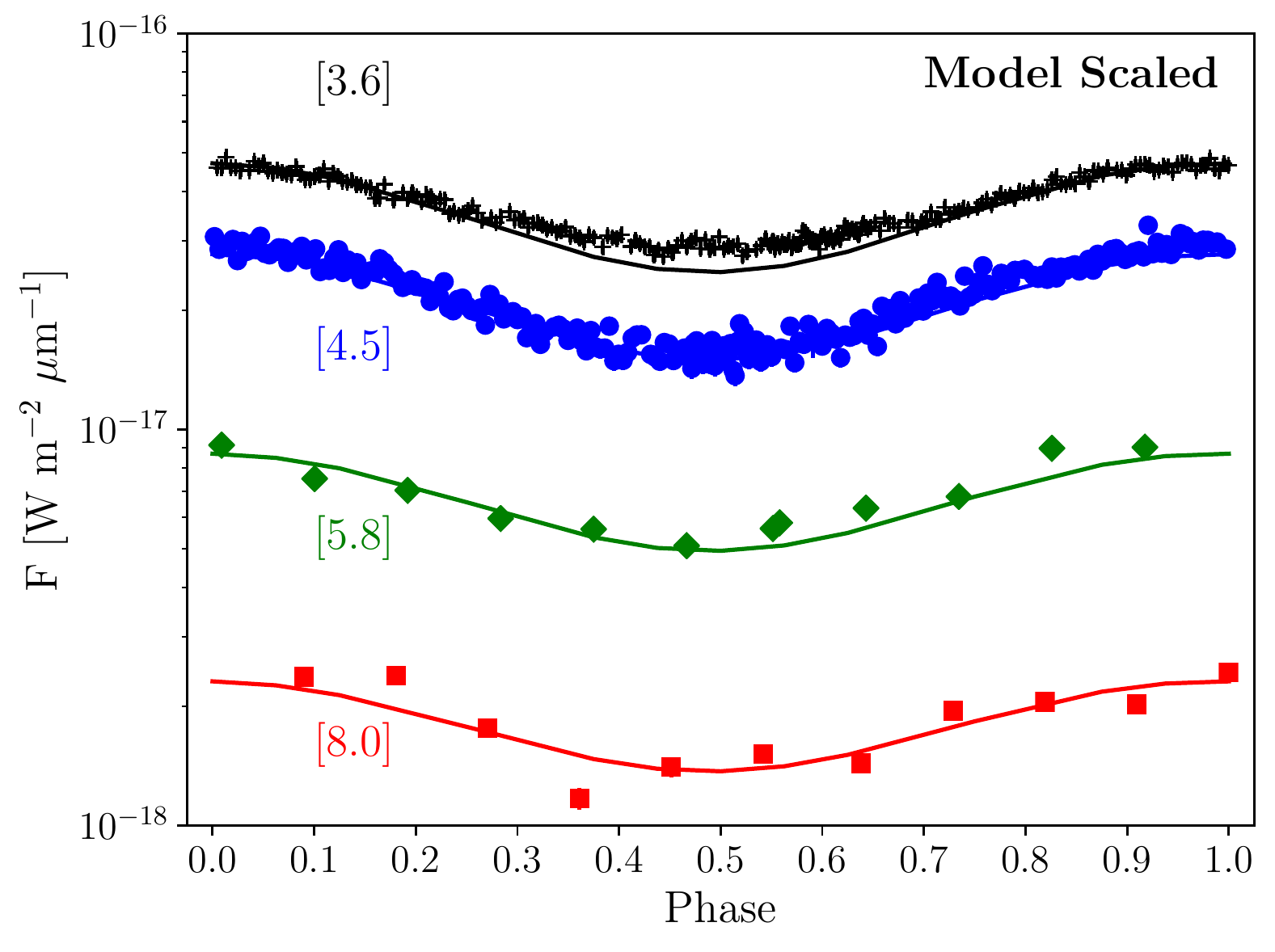}
   \caption{[3.6] (black), [4.5] (blue), [5.8] (green) and [8.0] (red) $\mu$m \textit{Spitzer} band phase curves from \citet{Casewell2015} (points) and post-processed GCM model fluxes (lines).
   The left plot shows the fiducial model fluxes while the right shows the model fluxes normalised to the average observed flux in each band, given by the relative factors in Table \ref{tab:shifts}.
   Errors in the observational data are on the order of the point size.}
   \label{fig:phase_Spitzer}
\end{figure*}

\begin{table}
\centering
\caption{Average relative phase curve flux shift required for the model in each IR band to match observed fluxes.}
\begin{tabular}{c c}  \hline \hline
 Band & Av. Rel. Shift   \\ \hline
 \textit{J} & 1.047  \\
 \textit{H} & 1.300  \\
 \textit{K$_{s}$} & 1.257  \\ \relax
 [3.6] & 2.282  \\ \relax
 [4.5] &  1.848 \\ \relax
 [5.8] &  2.662 \\ \relax
 [8.0] &  3.376 \\
\hline
\end{tabular}
\label{tab:shifts}
\end{table}

In this section, we produce synthetic phase curves of our model for the bands used in the \citet{Casewell2015} observational campaign.
We focus on the infrared photometric bands \textit{J}, \textit{H}, \textit{K$_{s}$} and \textit{Spitzer} 3.6, 4.5, 5.8 and 8.0$\mu$m bands.
For a fair comparison to the \citet{Casewell2015} data, the WD model flux in each band is added to each synthetic phase curve.
From Fig. \ref{fig:em_spec_2}, during nightside phases significant flux in the infrared, \textit{J}, \textit{H}, \textit{K$_{s}$} and \textit{Spitzer} bands is contributed by the WD.
Including this IR emission from the WD is required to `flatten out' the phase curves near 0.5 phase and better reproduces the observed phase curve shapes.

Our model generally over predicts the flux in each IR bandpass (LHS Fig. \ref{fig:phase_JHK} and \ref{fig:phase_Spitzer}) by factors of $\approx$1-3.
We scale each model phase curve by the relative difference between the average flux of the model and observations to compare the shape of the phase curve (RHS Fig. \ref{fig:phase_JHK} and \ref{fig:phase_Spitzer}).
Table \ref{tab:shifts} presents the required scaling factors for each band.
The scaled model phase curves match the observed phase curves shapes well, suggesting the peak to trough amplitudes are reasonably approximated in the GCM simulation.
The \textit{H}, \textit{K$_{s}$} and \textit{Spitzer} 3.6$\mu$m bands would require some additional flattening to better fit the observed shape.
This suggests that either the day-night contrast is too large in these bands (i.e. the day-night heat transport is too weakly modelled here), or the chemical abundances composition may be different to those calculated here, potentially through non-equilibrium effects.
No discernible phase curve offset is produced in the model output, with a highly symmetric profile, typical of current WD-BD phase curve data \citep[e.g.][]{Parsons2017, Casewell2018}.
This is due to the low inclination angle of 35$^{\circ}$ which does not have a large flux contribution from the westward shifted patters closer to the equator (Fig. \ref{fig:olr}).

\section{Discussion}
\label{sec:discussion}

In this section, we discuss our results in context and suggest additional considerations for future modelling efforts.

\subsection{Atmospheric structure and dynamics}

The main dynamical feature of the BD simulations is the significant narrower meridional extent of the equatorial super-rotating jet compared to typical HJ simulations \citep[e.g. see review by][]{Heng2015}.
Due to the fast rotation, high gravity (small scale height) of the BD, the Rossby deformation radius is significantly smaller compared to typical HJ systems.
This leads to the expected Matsuno-Gill flow pattern being compressed between $\sim$$\pm$30$^{\circ}$ latitude.

At higher latitudes, winds are much weaker than that at low latitudes, together with the fast rotation (large Coriolis force) implying a geostrophic circulation regime there.
As a result, wind vectors follow closely parallel to isotherms.
The horizontal thermal structure poleward of $\pm$45$^{\circ}$ latitude closely resembles an equilibrium structure.
This is consistent with the analytic wave solution of \citet{Showman2011a} assuming no frictional drag (which is effectively assuming geostrophy).
Finally, hourly outputs of our simulation also exhibit small scale instability mostly around mid latitudinal and counter-rotating flow regions, presumably caused by baroclinic instability.
These features are probably responsible for the dayside variability in the OLR (Sect. \ref{sec:OLR}).
Similar instabilities were examined by \citet{Showman2015, Fromang2016} and \citet{Menou2019} which focused on meandering of the equatorial jet in HJ simulations.

\citet{Carone2015, Penn2017, Penn2018} and \citet{Carone2019} investigate mechanisms for possible westward offsets in phase curves for tidally locked, fast rotating planets as the dynamics becomes more rotationally dominated.
As a highly rotational dominated regime, our simulations exhibit a similar mechanism to the above studies.
However, the Rossby wave gyres in our model are compressed closer to the equator, which also leads to a westward hot spot shift as viewed at equatorial latitudes.

In a parallel study, \citet{Tan2020} explored atmospheric circulation of tidally locked WD-BD systems with decreasing rotation period down to 2.5 hours.
The atmospheric circulation of the most rapidly rotating case is qualitatively similar to the results presented here, showing a narrowing equatorial super-rotating jet, nearly geostrophic flows at mid-high latitudes and larger day-night temperature difference than those of typical hot Jupiter simulations.
Although using different GCM and radiative forcing setup, the agreement between two studies is quite encouraging.
\citet{Carone2019}, \cite{Tan2020} and this study also show agreement on the formation of a westward hotspot shift as a consequence of the fast rotation rate.

Future studies on similar objects should consider increasing the resolution of the GCM simulation in order to capture the smaller scale features.
Tests performed in \citet{Showman2015} (however for slower rotation rates than modelled here) suggest a resolution of C48 is sufficient to capture the larger scale dynamical features.

We note that \citet{Carone2019} suggest that when simulating faster rotating hot Jupiter objects (P$_{\rm orb}$ $\lesssim$ 1.5 days ) that a deeper lower boundary (e.g. P$_{\rm 0}$ = 700 bar) is used.
In this study we chose a bottom boundary of 220 bar in order to directly compare to typical HJ GCM set ups.
Although we do not capture potentially important deeper atmospheric motions, our set-up captures the important photospheric pressures where most of the observable flux emerges from.
Future GCM modelling efforts for short period WD-BD should strongly consider simulating to a greater atmospheric pressure in order to capture any deeper dynamical phenomena, but also to more accurately simulate the radiative-transfer in the deeper optically thick regions, not probed in the current study.
However, a surface boundary of P$_{\rm 0}$ = 1000 bar would have an IR surface optical depth of $\tau_{\rm L_{eq}}$ $\approx$ 100 for our simulation, still an order of magnitude below typical HJ values \citep[e.g.][]{Heng2011b, Rauscher2012}.
At these high pressures H$_{2}$-H$_{2}$ and H$_{2}$-He continuum opacity is an important IR opacity source, and so a power-law dependence on $\tau_{\rm L_{eq}}$ \citep[e.g.][]{Heng2011b, Rauscher2012} may be more appropriate to include for a double-grey RT schemes.

\subsection{Vertical mixing and non-equilibrium chemistry}

\begin{figure*} 
   \centering
   \includegraphics[width=0.49\textwidth]{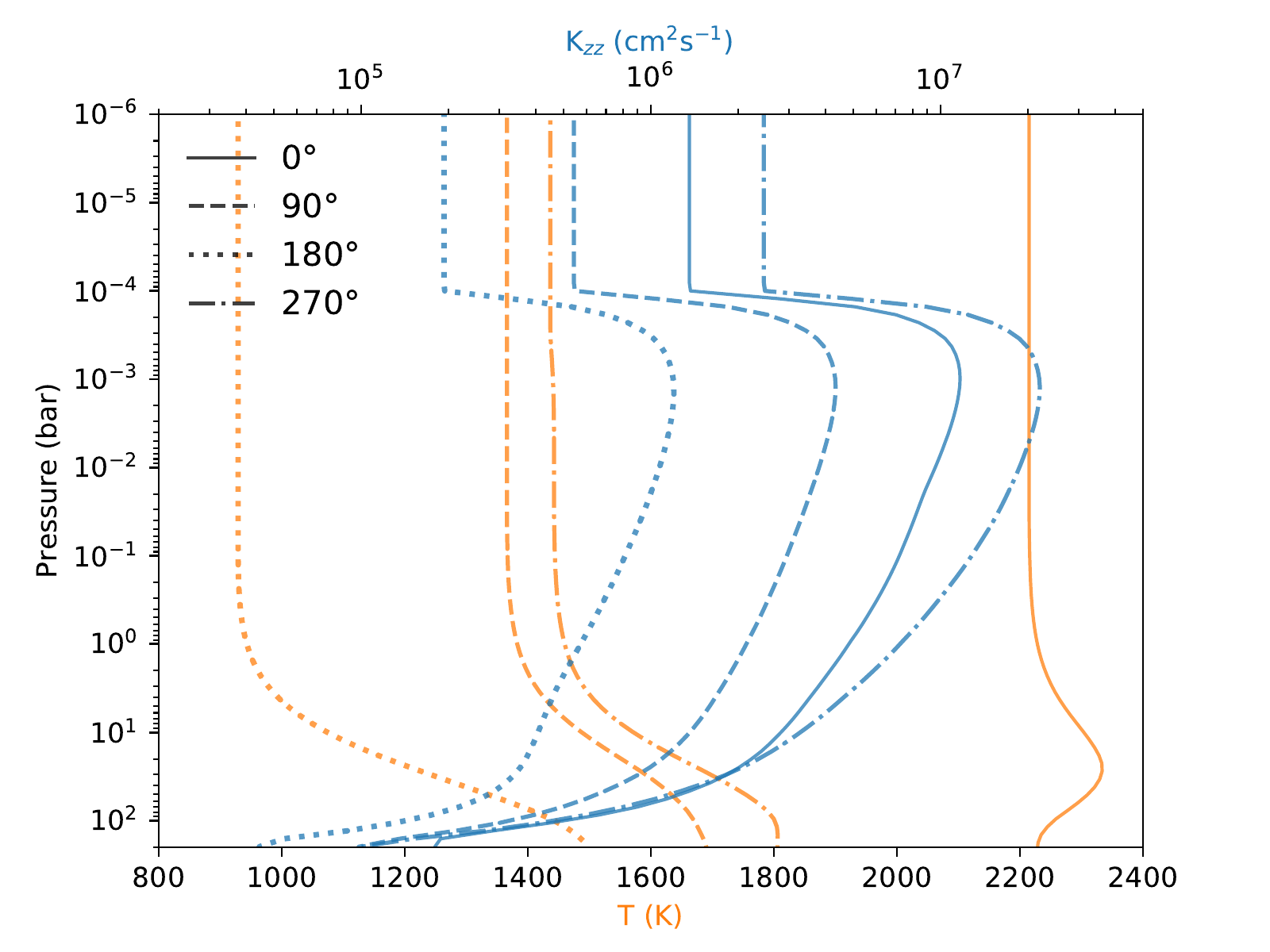}
   \includegraphics[width=0.49\textwidth]{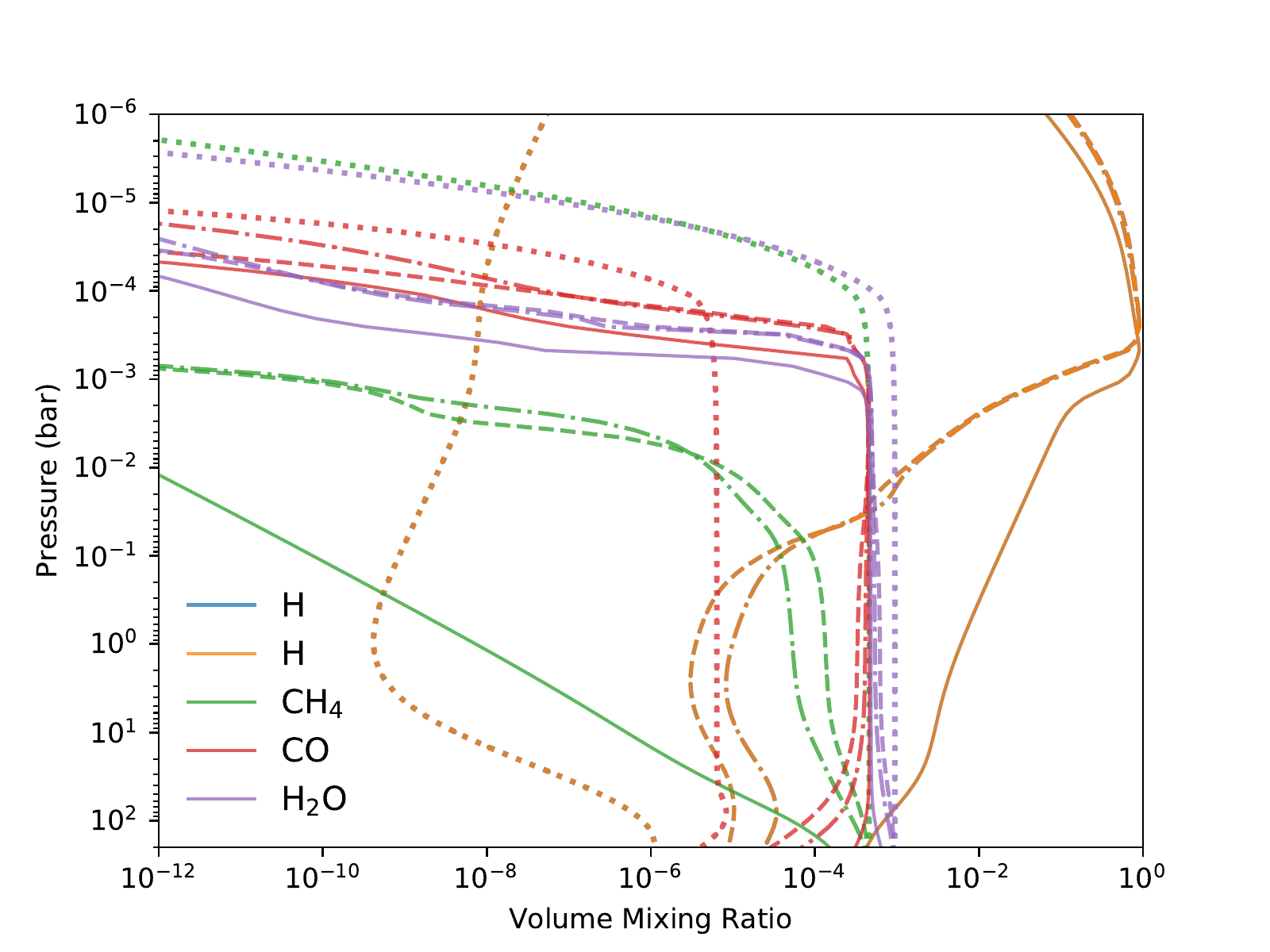}
   \includegraphics[width=0.49\textwidth]{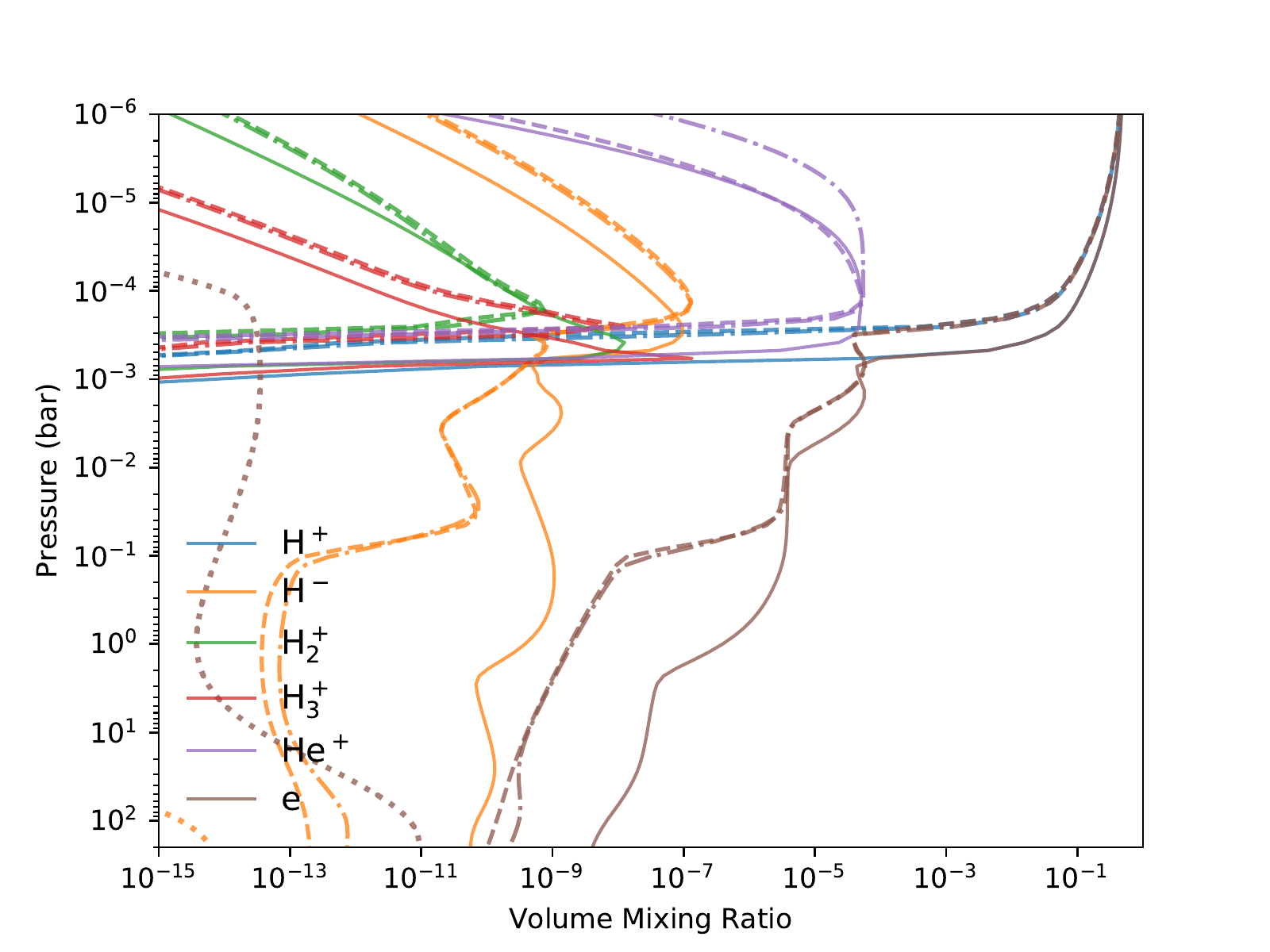}
   \includegraphics[width=0.49\textwidth]{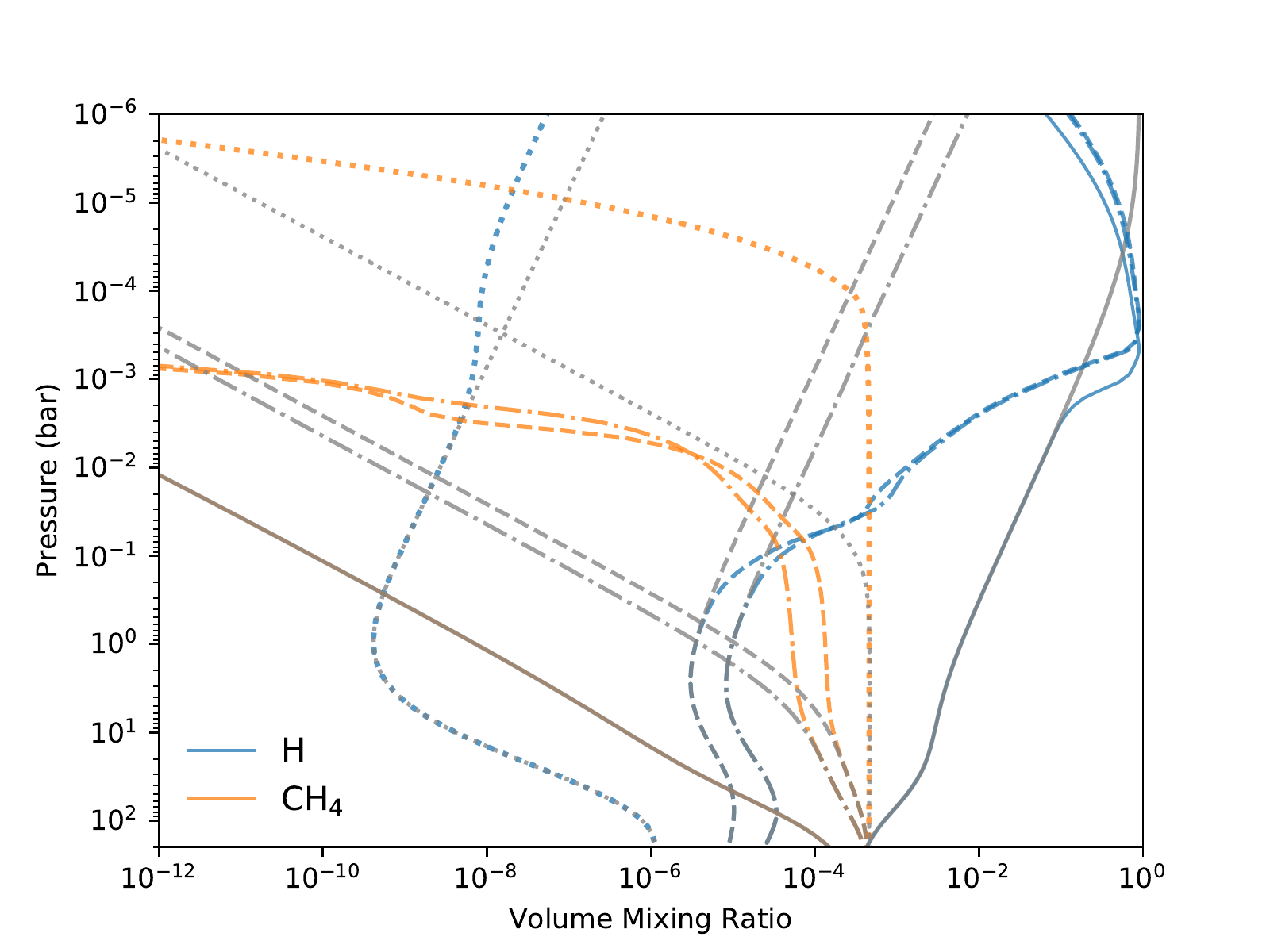}
   \caption{Top left: T-p profiles with the derived K$_{\rm zz}$ from the GCM model for each quadrant.
   Top right: \textsc{vulcan} results for the neutral species.
   Bottom left: \textsc{vulcan} results for the ion species.
   Bottom right: Chemical equilibrium values (grey) compared to the  \textsc{vulcan} results for CH$_{4}$ and H.
   Solid, dashed, dotted and dash-dotted lines correspond to the 0$^{\circ}$, 90$^{\circ}$, 180$^{\circ}$ and 270$^{\circ}$ quadrants respectively.}
   \label{fig:chem}
\end{figure*}

Recent observations of the shorter period HJ WASP-43b \citep{Chubb2020} and cool Brown dwarfs \citep{Miles2020} suggest non-equilibrium chemistry is an important consideration that shapes the spectrum of objects in a similar regime to WD0137-349B.
The weak vertical velocities and overturning seen in the GCM model at low-mid latitudes suggest that the upper atmosphere near the equatorial regions are slowly replenished from the deeper regions (Fig. \ref{fig:map_w}).
This may act to starve the supply of photochemically active species to the upper atmosphere.
To examine this we utilise the 1D version of the chemical kinetic model \textsc{vulcan} \citep{Tsai2017}.
We calculate a K$_{\rm zz}$ [cm$^{2}$s$^{-1}$] profile from the GCM results using the relation K$_{\rm zz}$ = H $\cdot$ w$_{\rm z}$ \citep[e.g.][]{Moses2011}, where H is the scale height and w$_{\rm z}$ the rms vertical velocity across the quadrant.
We average this value and the temperature-pressure profile across four quadrants, the dayside, nightside and east and west hemispheres.
Our profiles are also extended to 10$^{-8}$ bar to capture the important upper atmospheric regions for UV photo-chemistry.
These are then used as input to the \textsc{vulcan} model.
We include photochemical and ion chemistry in the kinetic model.
For the UV incident beam, we assume an hemispheric average zenith angle of 46$^{\circ}$ for the dayside profile and 74$^{\circ}$ for the terminator hemispheres.

Figure \ref{fig:chem} shows the results of the \textsc{vulcan} model for each quadrant.
It is clear that the overturning features on the western terminator (Fig. \ref{fig:map_w}) produce the largest K$_{\rm zz}$ profiles, but mixing is overall weak at K$_{\rm zz}$ $\sim$ 10$^{6}$ cm$^{2}$s$^{-1}$.
Our results show that significant photo-chemical induced ionisation of Hydrogen and photo-chemical disassociation of molecules occurs at the upper atmosphere (p $<$ 10$^{-4}$ bar) for the dayside and terminator hemispheres.
This region is dominated by a large fraction of neutral hydrogen, ionised hydrogen and free electrons.
The bottom right panel of Figure \ref{fig:chem} compares the CH$_{4}$ chemical equilibrium and \textsc{vulcan} results, suggesting that CH$_{4}$ is quenched from 10-100 bar on the east and west terminator regions, but in equilibrium on the dayside and nightside.

A 3D examination of the non-equilibrium chemistry could be considered for future studies, similar to recent GCM studies \citep[e.g.][]{Bordwell2018, Drummond2018b, Mendonca2018, Steinrueck2019, Drummond2020}.
Our results suggest consistent photo-chemical and ionisation effects in the 3D model is also likely to be more important in the WD-BD cases than the HJ cases due to the high amount of UV flux the BD receives.

\subsection{UV effects and heating in the upper atmosphere}

Our modelled emission spectra were not able to reproduce the elemental emission features from the BD reported in \citet{Longstaff2017}.
This suggests that other mechanisms beyond that modelled in the GCM are required to produce an upper atmospheric temperature inversion.
Absorption of UV photons and subsequent heating in the upper atmosphere is a possible candidate for producing a temperature inversion.
The energy release by recombination of photochemical products occurring in the upper atmosphere could also be a source of significant heating.
\citet{Longstaff2017} also suggest a chromospheric-like upper atmospheric region which thermally disassociates molecules in addition to providing a strong temperature inversion region.

In this study we have used the solar metallicity grey opacity parameters from \citet{Guillot2010}, tuned to reproduce an HD 209459b-like structure and irradiation by a main sequence star at optical wavelengths.
One of the major uncertainties for our modelling is the UV opacities for the BD atmosphere, and therefore the radiative heating from the primary UV irradiation by the WD.
The bond albedo is also a major uncertainty, should it be higher than that assumed here (A$_{\rm B}$ = 0.1), the atmosphere would be cooler than that modelled here.

We suggest a possible way to approximate the radiative heating from UV irradiation would be to split the shortwave scheme in the double-grey radiative-transfer into a ultra-violet and visible component, each with a fraction of the total irradiative flux and separate `Bond albedo'.
The reference optical depth of the UV band can then be tuned to match the expected $\tau$ $\sim$ 1 at lower pressures ($\approx$ $\mu$bar to mbar) informed by photochemical kinetics modelling \citep[e.g.][]{Lavvas2014, Rimmer2016} or 1D radiative-convective modelling \citep[e.g.][]{Lothringer2018}.
Absorption of more shortwave energy in the upper atmosphere would also change the dynamical structure of the atmosphere, producing a shallower dynamical layer, potentially increasing the variability (and phase offset) of the photospheric regions compared to the current study.
If the shortwave absorption becomes significant compared to the longwave absorption then the formation of a temperature inversion is also more favoured.
The energy released from chemical recombination by photochemical products can also estimated as a function of the UV band flux and the available photochemical products.
A simplified, net photochemical species passive tracer scheme could also be included in the GCM to more accurately inform the replenishment rates of photochemical products to the upper atmosphere.
A 3D chemical kinetics scheme was used in \citet{Yates2020} in a similar manner to model ozone production on Proxima Centauri b.
The above schemes will be experimented with in our future modelling efforts.

\subsection{Cloud formation and effect on OLR}

\begin{figure} 
   \centering
   \includegraphics[width=0.49\textwidth]{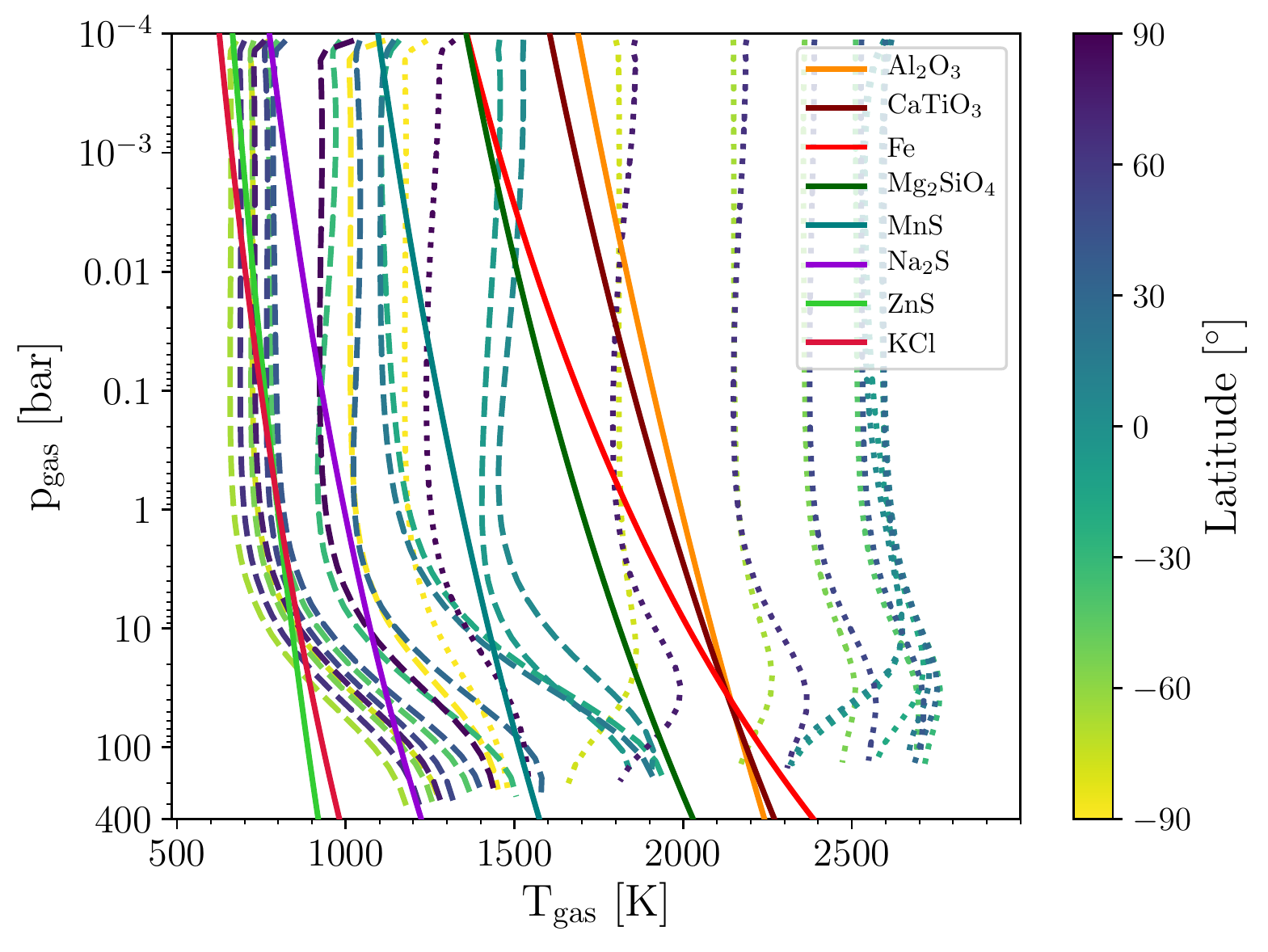}
   \caption{T-p profiles from -90$^{\circ}$ to 90$^{\circ}$ latitude (colour bar) at a longitude of 0$^{\circ}$ (dotted lines) and 180$^{\circ}$ (dashed lines).
   Coloured solid lines denote the supersaturation zones of various mineral species at solar metallicity \citep{Lodders2002,Visscher2006, Visscher2010, Morley2012, Wakeford2017a}.}
   \label{fig:gc_cloud}
\end{figure}

The temperature structure results of the GCM model suggest that mineral cloud formation is likely to occur.
Figure \ref{fig:gc_cloud} presents 1D T-p profiles from different locations from the GCM with the supersaturation curves of mineral materials from \citet{Lodders2002,Visscher2006, Visscher2010, Morley2012} and \citet{Wakeford2017a} at solar elemental ratios.
This plot suggests that significant cloud formation of multiple species is expected to occur on the nightside of the BD.
The refractory elements and silicates are expected to form at deeper pressures than simulated here, however, mineral sulphide and salt species are likely to form in the nightside photospheric regions.
We suggest the additional IR opacity provided by these clouds may act to warm the nightside regions by reducing the efficiency of atmospheric cooling.

Figure \ref{fig:gc_cloud} suggests refractory and silicate minerals can potentially form at higher latitudes on the dayside of the BD.
This may have a more direct impact on the emission spectra, and hence phase curves, by providing additional opacity to the upper parts of the atmosphere that contribute the most flux to the synthetic observations.
Much of the high latitude dayside regions are constantly in the line of sight at an orbital inclination of 35$^{\circ}$, so this opacity may act to reduce the outgoing IR flux, generally over predicted in our modelled phase curves (Fig. \ref{fig:phase_JHK} and \ref{fig:phase_Spitzer}).
The weak vertical velocities and strong gravity of the BD suggest that only small, sub-micron cloud particles would be able to remain lofted in the photospheric regions.

We note the specific cloud structure will also depend on the internal temperature of the planet.
Recent observations of WASP-121b by \citet{Sing2019} suggested Mg and Fe atoms present at high altitudes in the planet, potentially indicating a high internal temperature (T$_{\rm int}$ $\sim$ 500 K) which does not allow the condensation of refractory material at greater pressures.
High resolution spectra by \citet{Longstaff2017} of WD0137-349B show that the strength of the refractory elements (e.g. Mg, Fe, Si) decrease on nightside phases of the BD, indicating possible active condensation processes occurring in the atmosphere.

\section{Summary and Conclusions}
\label{sec:conclusions}

Short period White dwarf - Brown dwarf binary systems offer a unique opportunity to explore irradiated atmospheres under more extreme conditions than typical hot Jupiter systems.
In this study, we presented an initial exploration of the 3D atmospheric properties of the Brown dwarf WD0137-349B.
We utilised the Exo-FMS GCM model with a dual band grey radiative-transfer scheme to model the thermal and dynamical properties of the Brown dwarf atmosphere.
We used the 3D radiative-transfer model \textsc{cmcrt} to post-process the GCM output and produce synthetic emission spectra and phase curves.

Our modelling efforts suggest the atmosphere exhibits a combination of the dynamical properties expected from theory and previous (ultra) hot Jupiter studies, from the strong irradiation, high surface gravity and short rotation period of the Brown dwarf.
Our results are summarised as follows:
\begin{itemize}
\item A large day-night contrast is seen in the GCM as expected from theory.
\item Generally inefficient day-night energy transport, except near the equatorial jet region.
\item Generally weak vertical velocities with overturning structures on the western terminator regions.
\item Phase curve shapes are generally well fit, but the absolute flux is over-predicted by a factor of $\approx$1-3 dependent on the photometric band.
\item Photochemistry produces a significantly ionised upper atmospheric region.
\end{itemize}

Future modelling efforts can improve on the accuracy of our presented model with a few additions, for example
\begin{itemize}
\item Extending the simulation boundaries to the deeper, optically thick atmospheric regions.
\item Modelling the effect of UV photochemical products and radiative heating on the thermal structures in 3D.
\item Inclusion of a cloud formation and radiative feedback scheme.
\end{itemize}

Current and future photometric and spectroscopic instrumentation presents an exciting opportunity to observe WD0137-349B and other White dwarf - Brown dwarf short period binary systems in more precise detail.
Such data would help further constrain the unique atmospheric properties of objects in this dynamical and radiative parameter regime, and test the theory and modelling of these objects to widen a holistic understanding of irradiated atmospheres in general.

\section*{Acknowledgements}
We thank the reviewer for constructive advice and suggestions on the manuscript content.
We thank V. Parmentier for advice on dynamical regime scales.
G.K.H. Lee thanks J. Barstow for \textsc{nemesis} formatted k-tables and members of the UK exoplanet community for discussion and encouragement on this project.
G.K.H. Lee acknowledges support from the University of Oxford and CSH Bern through the Bernoulli fellowship.
S.L. Casewell acknowledges funding from the STFC Ernest Rutherford Fellowship program.
K.L. Chubb acknowledges funding from the European Union's Horizon 2020 Research and Innovation Programme, under Grant Agreement 776403.
M. Hammond acknowledges support through the STFC studentship program.
Plots were produced using the community open-source Python packages Matplotlib \citep{Hunter2007}, SciPy \citep{Jones2001}, and AstroPy \citep{Astropy2018}.
Our local HPC support at Oxford is highly acknowledged.
This work was supported by European Research Council Advanced Grant \textsc{exocondense} (\# 740963).




\bibliographystyle{mnras}
\bibliography{bib2} 



\appendix

\section{g-ordinate emission composite biasing}
\label{app:1}

In \citet{Lee2019} a method for using the correlated-k approximation was presented for computing emission spectra in a Monte Carlo radiative-transfer context.
This used an unbiased sampling for the g-ordinate of each photon packet, g$_{\rm samp}$, emitted in each cell, given by
\begin{equation}
g_{\rm samp} = \frac{w_{\rm g}L_{\rm g}}{\sum_{g}w_{\rm g}L_{\rm g}},
\end{equation}
where w$_{\rm g}$ is g-ordinate weight and L$_{\rm g}$ [erg s$^{-1}$ cm$^{-1}$] the luminosity contributed by that k-coefficient in a cell.
This scheme has the property that the higher numbered g-ordinate will usually be more likely to be sampled, since generally w$_{\rm g}$L$_{\rm g}$ $<$ w$_{\rm g+1}$L$_{\rm g+1}$, unless the opacity distribution is flat.
In some bands where the opacity distribution has a large gradient, for example near a line center, the lower g-ordinates may be under sampled, leading to unwanted noise by not sampling the true opacity distribution adequately.

To alleviate this we follow a composite biasing scheme similar to \citet{Baes2016} where the g-ordinate is sampled from a the unbiased probability distribution function, p(g), and a uniform distribution function, q(g), given by
\begin{equation}
q_{\star}(g) = (1 - \xi)p(g) + \xi q(g) = \frac{(1 - \xi)w_{\rm g}L_{\rm g}}{\sum_{\rm g}w_{\rm g}L_{\rm g}} + \frac{\xi}{N_{\rm g}},
\end{equation}
where N$_{\rm g}$ is the number of k-coefficients in the band, and $\xi$ = [0,1] the composite biasing factor.
The weight of the photon packet is then
\begin{equation}
W_{\rm ph} = \frac{1}{(1 - \xi) + \xi\langle L_{\rm g}\rangle / w_{\rm g}L_{\rm g}},
\end{equation}
where
\begin{equation}
\langle L_{\rm g}\rangle = \frac{\sum_{\rm g}w_{\rm g}L_{\rm g}}{N_{\rm g}}.
\end{equation}


\bsp	
\label{lastpage}
\end{document}